\documentstyle[twocolumn,prb,aps,eqsecnum]{revtex}

\input{epsf}

\begin{document}
\draft
\title{
Test of the Kolmogorov-Johnson-Mehl-Avrami 
picture of metastable decay in a model with 
microscopic dynamics
}

\author{Raphael A.\ Ramos*} 
\address{
Department of Physics, University of Puerto Rico, Mayaguez, Puerto Rico 00681
}
\address{
and 
Supercomputer Computations Research Institute
and Center for Materials Research and Technology, 
Florida State University, Tallahassee, Florida 32306-4130
}
\author{Per Arne Rikvold\dag} 
\address{
Center for Materials Research and Technology, 
Department of Physics, 
and 
Supercomputer Computations Research Institute,
Florida State University, Tallahassee, Florida 32306-4350
}
\address{
and 
Colorado Center for Chaos and Complexity,
University of Colorado, Boulder, Colorado 80309-0216
}
\address{
and 
Department of Fundamental Sciences,
Faculty of Integrated Human Studies,
Kyoto University, Kyoto 606, Japan
}
\author{M.~A.\ Novotny\ddag}
\address{
Supercomputer Computations Research Institute,
Florida State University, Tallahassee, Florida 32306-4130
}

\date{\today}

\maketitle

\begin{abstract}
The Kolmogorov-Johnson-Mehl-Avrami (KJMA) theory for the time evolution of
the order parameter in systems undergoing first-order phase transformations
has been extended by Sekimoto to the level of two-point correlation
functions. Here, this extended KJMA theory is applied to a kinetic Ising 
lattice-gas model, in which the elementary
kinetic processes act on microscopic length and time scales. The 
theoretical framework is used to analyze data from extensive 
Monte Carlo simulations. 
The theory is inherently a mesoscopic continuum picture, 
and in principle it requires a large
separation between the microscopic scales and the mesoscopic scales
characteristic of the evolving two-phase structure.
Nevertheless, we find excellent quantitative
agreement with the simulations in a large parameter regime, extending
remarkably far towards strong fields (large supersaturations) and
correspondingly small nucleation barriers. The original KJMA
theory permits direct measurement of the order parameter in the metastable
phase, and using the extension to correlation functions one can also
perform separate
measurements of the nucleation rate and the average velocity of the convoluted
interface between the metastable and stable phase regions. The
values obtained for all three quantities are verified by other 
theoretical and computational methods.
As these quantities are often difficult to measure directly during a
process of phase transformation, data analysis using the extended
KJMA theory may provide a useful experimental alternative.
\end{abstract}
\vskip2pc

\narrowtext

\section{Introduction}
\label{sec:1}

The phase-transformation kinetics in systems undergoing  
first-order phase transitions 
are important in many scientific and technological 
disciplines. Around 1940, Kolmogorov,\cite{KOLM37}
Johnson and Mehl,\cite{JOHN39} and Avrami\cite{AVRAMI} (KJMA)  
introduced a simple theory describing 
the decay of a metastable system towards a unique equilibrium phase. 
This theory applies to systems with a nonconserved order parameter, in 
which the decay is driven by a difference between 
the free-energy densities of the metastable and equilibrium phases. 
Originally conceived for metallurgical applications, it 
was later formalized and generalized by Evans,\cite{EVAN45} who also 
pointed out its applicability in surface science. 

The basic assumptions of the KJMA theory are simple: negligibly small  
``droplets'' of the equilibrium phase nucleate from the
uniform metastable phase and 
subsequently grow without substantial deformation. 
The growing droplets are assumed to be randomly placed and overlap freely, 
with the result that the remaining volume fraction occupied by the 
metastable phase decays exponentially with a power of time:
\begin{equation}
\label{eq:intro}
\varphi_{\rm ms}(t) 
= \exp \left( - C t^\alpha \right) \, . 
\end{equation}
Both the coefficient $C$ and the ``Avrami exponent'' $\alpha$ 
depend on the spatial 
dimension $d$ and on details of the nucleation and growth 
processes. In general, $\alpha > 1$; thus Eq.~(\ref{eq:intro}) does 
{\em not\/} represent a stretched exponential.

The KJMA picture has been extensively applied in diverse fields of 
research.\cite{ISI} 
A small selection of examples include 
transitions between different liquid-crystal phases;\cite{PRIC71} 
crystallization kinetics in 
lipids,\cite{LIPID} 
sugars,\cite{SUGAR} 
polymers,\cite{HUAN98} 
and eutectic mixtures;\cite{HORT93,ELDE94} 
solid-state phase 
transformations;\cite{YAMA84,AXE86,HERM97,RICK97,JOU97}
domain switching in 
ferroelectrics\cite{ISHI71,CHAN89,DUIK90,BEAL94,ORIH92,ORIHASH94,%
MITO94,RICI98,SHUR98} 
and 
ferromagnets;\cite{HIRS92,RICH94,RICH95C,RICH96,RIKV97B,KOU98,CHOE97}
adsorption and surface growth kinetics in 
electrochemical\cite{BOSCO,HOLZ94,MAES94,SARA94,DRET97,FUKU97,%
RIKV97,RIKV98,FANE98,FINN98}
or gas/vacuum environments;\cite{BECK92} 
rock formation;\cite{HORT93,GEOLOGY} 
and slow combustion.\cite{KART98} 

During the first forty years after its inception, the KJMA theory 
appears to have been appreciated mostly by experimentalists. 
It attracted little sustained theoretical interest 
until the 1980s, when Sekimoto derived 
exact expressions for the two-point phase correlation function within the KJMA 
picture.\cite{SEKIMOTO}
Knowledge of correlation functions makes it possible to predict and 
interpret the results of small-angle scattering experiments. 
In this paper we demonstrate how it can also be utilized to obtain 
independent estimates of the droplet nucleation rate and growth velocity. 
Sekimoto's results have been generalized 
to systems with infinitely degenerate equilibrium phases\cite{AXE86} 
and to the case of finite degeneracy.\cite{OHTA87} 
We shall refer collectively to the theories 
that extend the KJMA picture to include correlation functions 
as the ``extended KJMA theory.'' 
Among recent applications of the extended KJMA theory are 
theoretical studies of domain switching in  
ferroelectrics\cite{CHAN89,DUIK90,BEAL94,ORIH92,ORIHASH94,%
MITO94,RICI98,SHUR98} 
predictions of magnetic-force-microscopy observables for nanoscale 
ferromagnets,\cite{RICH94,RICH95C,RICH96,RIKV97B}
and studies of hysteresis in spatially extended 
systems.\cite{ORIHASH94,MITO94,SIDES1,SIDES2}
Relations between two-point correlation functions and 
droplet size distributions have also been discussed recently.\cite{GARR97} 

There are many simplifying assumptions inherent in the extended KJMA 
theory. These include a constant nucleation rate, a constant interfacial 
propagation velocity, and the neglect of surface-tension effects that one 
would expect to become important when growing droplets meet and coalesce. 
In view of this, it is remarkable that the theory appears to perform as 
well as it does for such a wide variety of systems. 
Computer studies have previously been performed for models 
in which droplets of a fixed shape nucleate, either 
deterministically or randomly, and thereafter grow 
deterministically.\cite{RICK97,JOU97,MEYE94,SESS96} 
However, we are only aware of a single previous study\cite{SHNE98} 
in which the kinetic processes 
act on length and time scales that are {\it microscopic\/} 
compared to the growing droplets, as one 
would expect for real physical and chemical systems. 
Such studies are desirable, as they permit direct verification of 
the regimes of validity for the individual assumptions 
as well as the consequences of 
particular assumptions not being exactly fulfilled. 
In the present paper we further address this need, with particular emphasis 
on the information which can be obtained by combining results for the 
volume fraction and the two-point correlation functions. 
Our purpose is two-fold, as discussed below. 

Our first aim is to perform a detailed test of the validity of the 
extended KJMA theory in a particular model system with microscopic kinetics. 
We consider a two-dimensional 
kinetic Ising model of ferromagnetic or ferroelectric switching. 
This is equivalent to a simple lattice-gas model of submonolayer 
chemisorption onto a single-crystal surface in the limit that 
lateral diffusion can be ignored.\cite{RIKV97} For this model  
we test the KJMA predictions for time-dependent quantities that describe 
the mesoscopic two-phase 
structure during the evolution towards equilibrium. 
These quantities include the volume fraction, 
the two-point correlation function, and its Fourier transform, the structure 
factor. We find excellent quantitative
agreement between the theoretical predictions and the 
simulation results for a considerable range of applied fields and times. 

Our second aim springs from the significant 
regime of agreement that we establish between the KJMA predictions 
and the simulation results for the model studied. This 
enables us to use the KJMA prediction for the volume fraction 
to subtract those regions of the system which have already decayed to the 
stable phase at a particular time, and thus measure extended time and 
volume averages of the quasi-equilibrium order parameter in the metastable 
phase. This approach represents a novel method to measure ``thermodynamic'' 
quantities in a constrained metastable ensemble.\cite{RIKV94}
Furthermore, the extended KJMA predictions for the two-point correlation 
functions enable us to measure separately the nucleation rate and 
the average propagation velocity of the convoluted, moving 
interface between the metastable and stable phase regions. 
We show that the measured quantities agree well with 
theoretical predictions obtained by different methods. 
These methods are numerical transfer-matrix calculations for the metastable 
order parameter,\cite{NEWM77,MCCR78,PRIVMAN,CCAG93,CCAG94A}
a field-theoretical result for the nucleation 
rate,\cite{LANGER,GNW80}
and a nonlinear response theory for the interface velocity.\cite{RIKV99} 
The close agreement between the KJMA and independent estimates confirms that 
the KJMA theory does not merely produce good fits to the simulation 
data, but that the fitting parameters measure nonequilibrium 
physical quantities that can be difficult to measure by other methods. 

The remainder of this paper is organized as follows. 
In Sec.~\ref{sec:2} 
we introduce the model and the numerical methods used in this work. 
In Sec.~\ref{sec:2X} we summarize those results of 
the nucleation theory of metastable decay and the extended KJMA theory 
that are relevant to our study. We give most of the theoretical results for 
general spatial dimension $d$. 
In Sec.~\ref{sec:3} our two-dimensional 
simulation results are presented and explained in the framework 
of the theory. The regimes of agreement between theory and simulations 
are identified, and the simulation data are used to measure the 
metastable order parameter (Sec.~\ref{subsec:3.1}), 
as well as the average interface velocity and the nucleation rate 
(Sec.~\ref{subsec:4.2}), all as functions of the applied field.
Correlation functions (Sec.~\ref{subsec:4.3}) 
and structure factors (Sec.~\ref{subsec:4.5}) are also measured. 
The measured quantities are compared with independent theoretical estimates. 
In Sec.~\ref{sec:4} we summarize our results and give our conclusions 
and some suggestions for further study.

\section{Model and Methods}
\label{sec:2}

\subsection{Model Hamiltonian}
\label{subsec:2.1a}

The discrete model studied here is defined by the standard Ising Hamiltonian 
(the transformation to lattice-gas language is given below),
\begin{equation}
{\cal H} = -J \sum_{\langle i,j \rangle} s_{i}(t) \; 
s_j(t) - H \sum_{i} s_{i}(t) \; ,
\label{2.1.1}
\end{equation}
where $s_{i}(t) = \pm 1$ is the 
``spin'' at lattice site $\vec{r}_{i}$ at time $t$, 
$J > 0$ is the
ferromagnetic coupling constant, and $H$ is the external field. 
The sums $\sum_{\langle i,j \rangle}$ and $\sum_{i}$ run over all 
nearest-neighbor pairs and over all sites on a $d$-dimensional 
hypercubic lattice of linear size $L$, respectively. 
The lattice constant defines our unit of length. 
While our theoretical discussion is for general $d$, 
all the simulations presented are for $d$=2. 
Three-dimensional systems will be considered in a forthcoming 
paper.\cite{TOWN98} 
In order to avoid complications due to 
boundaries, we use periodic boundary conditions throughout. 
(For recent discussions of boundary effects in metastable decay, 
see Refs.~\onlinecite{RICH96,CIRI98}.) 

The magnetization per unit cell is 
\begin{equation}
m(t) = \frac{1}{L^d} \sum_{i} s_{i}(t) \;.
\label{2.1.1a}
\end{equation}
Under equilibrium conditions $m$ is the order parameter conjugate to $H$. 
{}For $H$=0 and temperatures below a finite critical temperature $T_{\rm c}$, 
the model has two degenerate equilibrium 
phases in which the magnetization has a constant spontaneous 
magnitude $m_{\rm sp}(T) \! = \! m_{\rm s}(T,H$=0). 
In the presence of a nonzero 
field the phase degeneracy is lifted, and the stable equilibrium magnetization 
$m_{\rm s}(T,H)$ 
has the same sign as $H$. 

Although it is not absolutely essential for our 
study that $T_{\rm c}$ and $m_{\rm sp}(T)$ are exactly 
known for the two-dimensional square-lattice Ising 
model,\cite{ONSA44,YANG52} these and other exact results 
(see Sec.~\ref{subsec:nuc}) enable us to quantitatively compare 
the various parameter estimates obtained from our numerical simulations 
with independent theoretical estimates. 

The Ising formulation is conveniently symmetric under 
simultaneous reversal of $H$ and $s_{i}(t)$, and it can be directly 
applied as a simple model for highly anisotropic ferromagnetic 
and ferroelectric systems. 
A less explicitly symmetric formulation is particularly 
convenient for discussing crystallization and adsorption phenomena. 
It is the equivalent two-state, attractive lattice-gas
model.\cite{YANG52B,PATH96} 
In terms of the time-dependent, local occupation variables 
$c_{i}(t) \in \{ 0,1 \}$,  Eq.~(\ref{2.1.1}) takes the form
\begin{equation}
{\cal H} = - \Phi \sum_{\langle i, j \rangle} 
c_{i}(t) \, c_{j}(t) - \mu \sum_{i} c_{i}(t) 
+ \frac{L^d}{2} \left ( \mu - \frac{1}{2} \mu_{0} \right ) \;.
\label{2.1.1aa}
\end{equation}
The quantities appearing in the equivalent formulations of the
Hamiltonian, Eqs.~(\ref{2.1.1}) and~(\ref{2.1.1aa}), are linked by the
transformations
\begin{mathletters}
\begin{eqnarray}
c_{i}(t) & = & \frac{1}{2}
\left [ 1 + s_{i}(t) \right ] \;, \\
\Phi & = & 4 J \;, \\
\mu & = & 2H + \mu_{0} \; .
\label{2.1.1aaa}
\end{eqnarray}
\end{mathletters}
Here $\Phi$ is the attractive lattice-gas interaction energy, and $\mu$
is the chemical or electrochemical 
potential, whose value at coexistence (i.e., for
$H$=0) is $\mu_{0} \! = \! -2zJ \! = \! -z \Phi /2$, 
where $z$ is the coordination number 
($z \! = \! 2d$ for hypercubic lattices). 
The chemical potential is related to the (osmotic) pressure $p$ as 
$\mu - \mu_{0} = k_{B} T \ln (p/p_{0})$, where $p_{0}$ is the pressure
at coexistence, $T$ is the temperature, and $k_{\rm B}$ is Boltzmann's 
constant. The order parameter conjugate to $\mu$ is the
density (for $d$=2: the coverage),
\begin{equation}
\theta(t) = \frac{1}{L^d} \sum_{i} c_{i}(t) = \frac{1}{2}[1 + m(t)] 
\;.
\label{2.1.1b} 
\end{equation}

\subsection{Stochastic Dynamics}
\label{subsec:2.1b}

The Ising lattice-gas Hamiltonian 
does not impose a particular dynamic on the system. 
To study the approach to equilibrium 
under the influence of thermal fluctuations, we use 
the stochastic Glauber dynamic.\cite{GLAU63}
This dynamic is defined by the acceptance probability for a proposed flip 
of $s_i$, 
\begin{equation}
W  \left [ s_{i}  \rightarrow -s_{i} \right ] 
= \frac{\exp(-\beta \Delta E)}
{1+\exp(-\beta \Delta E)} \; ,
\label{2.1.2}
\end{equation}
where $\Delta E$ is the energy change associated with the attempted 
spin flip, and $\beta = 1/ k_{\rm B}T$.
The dynamic was implemented by the standard algorithm used for kinetic 
studies:\cite{BIND92B} a series of steps in 
which sites are chosen at random and flipped with 
probability given by Eq.~(\ref{2.1.2}). 

The Glauber-Ising model described above (or the same 
Hamiltonian with the closely related Metropolis dynamic\cite{METR53}) 
has previously been used to study a large number of phase-ordering phenomena 
in condensed-matter physics and other fields. 
The two-dimensional version corresponding to the model for which we 
present numerical results, has been applied to 
thermally activated switching in uniaxial 
ferroelectric\cite{ISHI71,CHAN89,DUIK90,BEAL94,ORIH92,ORIHASH94,%
MITO94,RICI98,SHUR98} 
and 
ferromagnetic\cite{RICH94,RICH95C,RICH96,RIKV97B,KOU98,CHOE97} 
thin films. 
The equivalent lattice-gas model should give a reasonable 
representation of the kinetics of submonolayer adsorption in 
chemical\cite{BOSCO,HOLZ94,MAES94,SARA94,DRET97,FUKU97,%
RIKV97,RIKV98,FANE98,FINN98}
and physical\cite{BECK92} systems in which lateral adsorbate 
diffusion can be ignored. 
Electrochemical underpotential deposition,\cite{HOLZ94} 
in which second-layer formation is heavily suppressed, might be 
approximated by this model,\cite{RIKV97} although more detailed 
agreement with experimental studies of dynamical phenomena\cite{HOLZ94} 
requires the inclusion of lateral diffusion.\cite{RIKV98,BROWN98,BROWN98B}

\subsection{Quantities Characterizing the Decay}
\label{subsec:2.1c}

We study the decay of a metastable phase by first preparing the system 
in an initial state of magnetization $m(0)$=$+1$. We then apply a constant 
magnetic field $H < 0$ and let the system evolve in time. 
The decay of the metastable phase is characterized by the behavior of the 
relaxation function,\cite{BIND73A} which is defined in terms of $m(t)$ as
\begin{equation}
\phi(t) = \frac{m(t)-m_{\rm s}}{m(0)-m_{\rm s}} \; ,
\label{2.1.3}
\end{equation}
where $m_{\rm s}(T,H)$ is the 
field-dependent equilibrium magnetization. For 
$m_{\rm s} \! = \! -1$, $\phi(t)$ 
simply equals the time-dependent density $\theta(t)$ defined in 
Eq.~(\ref{2.1.1b}). The lifetime $\tau$ of the metastable phase 
is defined as the average first-passage time to $m$=0. 
In the parameter regime described by the KJMA theory, this is 
equivalent to the requirement 
that the ensemble average $\langle m(\tau)\rangle$=0. 
(For a discussion of the effects of using different cutoff values of $m$ 
to define $\tau$, see Ref.~\onlinecite{RIKV94A}.) 

During the decay process, we also 
compute the circularly averaged time-dependent 
structure factor $S(q,t)$ and correlation function $G(r,t)$.
The time-dependent structure factor $S(\vec{q},t)$ is 
defined in terms of the Fourier transform of $s_{i}(t)$,
\begin{equation}
\widehat{s}_{\vec{q}}(t) = \frac{1}{\sqrt{L^d}} 
\sum_{i=1}^{L^d} s_{i}(t) \; 
e^{- i\vec{q} \cdot \vec{r}_{i}} \;,
\label{2.1.5}
\end{equation}
as 
\begin{equation}
S(\vec{q},t) = \langle \widehat{s}_{\vec{q}}(t) \; 
\widehat{s}_{-\vec{q}}(t) \rangle - \langle m(t) \rangle^{2}
\delta_{\vec{q},\vec{0}} \; .
\label{2.1.4}
\end{equation}
The components of the reciprocal lattice vectors $\vec{q}$ 
are $q_{j\alpha} = 2 \pi j/L$ [$j = 0, \pm 1, \pm 2, \ldots, \pm (L/2-1), 
L/2$; $\alpha = x, y, \dots$], 
and $\delta_{\vec{q},\vec{0}}$ is the Kronecker delta function. 
The brackets $\langle\;\rangle$ imply 
an ensemble average over independent simulation runs.

The time-dependent two-point correlation function $G(\vec{r},t)$ is defined by
\begin{equation}
G(\vec{r},t) = \langle s_{i}(t)\;s_{j}(t)\rangle 
- \langle m(t) \rangle^{2} \; ,
\label{2.1.5a}
\end{equation}
where $\vec{r}$=$\vec{r}_{i}-\vec{r}_{j}$. It is circularly symmetric and 
was obtained as the inverse Fourier transform of Eq.~(\ref{2.1.4}). 
All Fourier transforms were computed with  
the Fast Fourier Transform subroutine {\tt fourn}.\cite{NUMRECC}
With the normalizations used here, 
$S(\vec{q} \! = \! 0,t) = L^d \, \mbox{Var}[m(t)]$, 
which is independent of $L$ if $G(\vec{r},t)$ is of finite range. 
Here, Var$[m(t)]$ is the time-dependent variance of the system magnetization, 
evaluated over an ensemble of independent runs. 
The quantities 
$S(\vec{q},t)$ and $G(\vec{r},t)$ were circularly averaged to obtain 
$S(q,t)$ and $G(r,t)$, respectively.

In order to find a time-dependent characteristic length scale, the first 
moment of $G(r,t)$ was computed using the definition
\begin{equation}
\langle r(t) \rangle = \frac{\sum_{r} r \; G(r,t)}
{\sum_{r} G(r,t)}.
\label{2.1.6}
\end{equation}

The temperature in all our simulations was fixed at 
$T \! = \! 0.8T_{\rm c}$, which is sufficiently far below $T_{\rm c}$ 
that the thermal correlation lengths in both the stable and metastable 
phases are microscopic and the main contributions to the correlation 
function come from the random two-phase structure. 
Other details about the simulations are given 
at the beginning of Sec.~\ref{sec:3}. 

\section{Theoretical Background}
\label{sec:2X}

\subsection{Nucleation Theory of Metastable Decay}
\label{subsec:nuc}

In order to obtain the characteristic times and lengths that are important 
to our analysis, we 
here give a brief summary of those aspects of the 
droplet theory of nucleation that are necessary to analyze our numerical 
results. More complete discussions can be found in 
Refs.~\onlinecite{RIKV94,RIKV94A,GUNT83B}. 

Thermal fluctuations in the uniform metastable phase create droplets of 
the stable phase. In terms of the droplet radius $R$, the free-energy 
difference between a uniform metastable system and one that contains a 
single such droplet is 
\begin{equation}
\label{eq:drfe}
\Delta F(R) = d \Omega_d^{\frac{d-1}{d}} R^{d-1} \sigma_0 - 
\Omega_d R^d |H| \cdot |m_{\rm s} - m_{\rm ms}| \,, 
\end{equation}
where $\Omega_d R^d$ is the volume of a droplet of radius $R$, 
$\sigma_0$ is the surface tension along a primitive lattice vector, 
and $m_{\rm ms}$ is the magnetization of the uniform metastable phase. 
The critical radius 
\begin{equation}
\label{eq:rc}
R_c = \frac{(d-1) \sigma_0}{|H| \cdot |m_{\rm s} - m_{\rm ms}|} 
\approx \frac{(d-1) \sigma_0}{2 |H| m_{\rm sp}} 
\end{equation}
corresponds to the maximum 
of $\Delta F(R)$. Droplets with $R \! < \! R_c$ almost always decay, whereas 
droplets with $R \! > \! R_c$ almost always continue to grow. 
In this work we only consider sufficiently strong fields that 
$R_c \! \ll \! L$. 
(The regimes of very weak fields, where this does not hold, are discussed in 
Refs.~\onlinecite{RICH94,RICH95C,RICH96,RIKV94,RIKV94A,JLEE94A}.) 
The nucleation rate per unit volume for growing regions of 
the equilibrium phase is determined by the free-energy cost of a critical 
droplet, $\Delta F(R_c)$, through a Van't Hoff-Arrhenius 
relation. The final form is shown by field-theoretical 
arguments\cite{RIKV94,LANGER,GNW80} 
to be 
\begin{mathletters}
\label{eq:I}
\begin{equation}
\label{eq:Ia}
I(T,|H|) 
\approx 
B(T) |H|^K \exp \left( - \frac{\Xi(T) + O(H^2)}{|H|^{d-1}} \right) \,, 
\end{equation}
where 
\begin{eqnarray}
\label{eq:Ib}
\Xi(T) &=& 
       \frac{\Omega_d \sigma_0^d }{k_{\rm B}T}
       \left( \frac{d-1}{|m_{\rm s} - m_{\rm ms}|} \right)^{d-1} \nonumber\\
       &\approx& 
       \frac{\Omega_d \sigma_0^d }{k_{\rm B}T}
       \left( \frac{d-1}{2 m_{\rm sp}} \right)^{d-1} . 
\end{eqnarray}
\end{mathletters}
In Eq.~(\ref{eq:Ia}), $B(T)$ is a nonuniversal prefactor. 
There is strong 
analytical\cite{LANGER,GNW80}
and numerical\cite{CCAG93,CCAG94A,RIKV94A,HARR84} evidence that the exponent 
$K$ is 3 for the two-dimensional Ising model, and it is 
believed to be $-1/3$ for the three-dimensional Ising model.\cite{GNW80}  
The approximate form of $\Xi(T)$ in Eq.~(\ref{eq:Ib}) is completely defined by 
quantities that for the two-dimensional Ising model are either known 
exactly ($\sigma_0$ and $m_{\rm sp}$),\cite{ONSA44,YANG52} 
or can be obtained through a Wulff construction 
by numerical integration of exactly known quantities 
($\Omega_d$).\cite{CCAG93,CCAG94A,ROTT81,ZIA82} 
The $O(H^2)$ corrections in Eq.~(\ref{eq:Ia}) are relatively 
minor\cite{RICH94} and will be ignored. 
 
The above calculations are based on the continuum assumption that 
$2 R_c \gg 1$. From Eq.~(\ref{eq:rc}) one sees that this requires that 
\begin{equation}
\label{eq:HMFSP}
H \ll (d-1) \sigma_0(T) / m_{\rm sp}(T) = H_{\rm MFSP}(T) \,. 
\end{equation}
This crossover field has been called {\em the mean-field spinodal\/} (MFSP), 
and the regime of stronger fields 
is called the {\em strong-field\/} regime.\cite{RIKV94A,TOMI92A}
Note that $H_{\rm MFSP}$ depends on $T$, but not on $L$.

\subsection{Continuum KJMA Theory}
\label{subsec:2.2}

\subsubsection{One-point averages}
\label{subsec:2.2a}

The KJMA theory of metastable decay 
describes the process of nucleation and growth in a large continuum 
system with a nonconserved order parameter and a nondegenerate equilibrium 
phase. (The precise meaning of ``large'' will be elucidated below.) 
It is assumed that the system is initially in a uniform 
metastable phase, in which critical droplets of the stable phase nucleate with 
rate $I(t)$ per unit volume and grow with radial growth velocity 
$v(t)$ without substantially altering their shapes. The two-phase 
structure of the system is represented by a phase field or indicator function, 
\begin{eqnarray}
u(\vec{r},t) = \left \{ \begin{array}{ll}
                     1 & \mbox{if $\vec{r}$ is in the 
                               metastable phase at $t$} \\
                     0 & \mbox{otherwise} \;.
                         \end{array}
               \right.
\label{2.2.1}
\end{eqnarray}
The simplest statistical quantity describing the structure of the system is 
the volume fraction of metastable phase, $\varphi_{\rm ms}(t)$, which is 
defined in terms of $u(\vec{r},t)$ by
\begin{eqnarray}
\varphi_{\rm ms}(t) = \langle u(t) \rangle = \frac{1}{L^d} \int 
\langle u(\vec{r},t) \rangle \; {d}^{d}{r} \;.
\label{2.2.2}
\end{eqnarray}
Assuming that droplets of stable 
phase nucleate with constant nucleation rate $I$ 
(the case discussed in Sec.~\ref{subsec:nuc}) 
and grow from an initial volume of zero ($R_c \ll L$) 
without interacting and with constant radial growth velocity $v(t) = v$, 
the ensemble average of $\varphi_{\rm ms}(t)$ is given by
\begin{eqnarray}
\varphi_{\rm ms}(t) 
&=& \exp \left[ - I \Omega_d v^d \int_0^t (t-s)^d {d} s \right] \nonumber\\
&=& \exp \left[ - \frac{\Omega_d I v^{d} t^{d+1}}{d+1} \right] \; ,
\label{2.2.3}
\end{eqnarray} 
an expression often referred to as 
``Avrami's law.''\cite{KOLM37,JOHN39,AVRAMI} The argument of the exponential 
(which grows without bound with time) 
is the ``extended volume'' of 
stable phase, obtained by adding the volume fractions of all droplets 
without correcting for overlaps. 
The exponential dependence on the extended volume is a special case
of a general result for randomly placed, polydisperse 
objects.\cite{POLYDISP} 
In the approximation considered here the distribution 
of droplet radii is uniform between $R_c \! \approx \! 0$ and $vt$. 
In response to claims that it does not represent a correct 
solution for the stochastic process defined in 
Refs.~\onlinecite{KOLM37,JOHN39,AVRAMI}, Eq.~(\ref{2.2.3}) was recently 
rederived without reference to the the extended 
volume.\cite{NOVEX} 
In Sec.~\ref{subsec:2.2c}  Eq.~(\ref{2.2.3}) will be used to provide 
explicit forms of the parameters in Eq.~(\ref{eq:intro}). 
Generalizations of Eq.~(\ref{2.2.3}), which consider complications such as  
finite-size effects, homogeneous and heterogeneous nucleation, anisotropic 
growth, and diffusion, are discussed in 
Refs.~\onlinecite{HERM97,RICI98,SHUR98,SARA94,YU,WEIN97}. 
Effects of nonstationary nucleation rates and growth velocities that depend 
on the droplet size are discussed in 
Refs~\onlinecite{SHNE98,WEIN97,SHNE93X}. 

Equation~(\ref{2.2.3}) defines the time scale 
$t_0 \! = \! A \left( I v^d \right)^{-\frac{1}{d+1}}$, in which 
$A = \left[ (d+1) \ln 2 / \Omega_d \right]^{\frac{1}{d+1}}$ depends 
weakly on $T$ through $\Omega_d$.  
This time approximately equals the metastable lifetime $\tau$ defined 
after Eq.~(\ref{2.1.3}). An important length scale is obtained from 
$t_0$ and the growth velocity $v$: 
\begin{equation}
R_0  =  v t_0  
= A v  \left(I v^d \right)^{- \frac{1}{d+1}} 
\approx v \tau \;. 
\label{eq:R0}
\end{equation}
This characteristic length describes the mesoscopic structure 
of the decaying system. 
It gives the average diameter of a droplet at 
$t \! \approx \! \tau $ and can
be seen as the average distance between independent droplets. 
The average number of droplets that contribute to the decay
is proportional to $(L/R_0)^d$. For Eq.~(\ref{2.2.3}) to describe the time 
evolution correctly, the system must contain a large number 
of independently nucleating and growing droplets, {\it i.e.\/},   
$(L/R_0)^d \! \gg \! 1$. This is the sense in which the system must be large. 
Because of the large number of droplets that contribute to the 
growth, the regime in 
which KJMA theory is expected to be valid is called the {\it multidroplet\/} 
(MD) regime.\cite{RIKV94A,TOMI92A}
Under these conditions the system is
self-averaging\cite{MILC86} and behaves approximately
deterministically according to Eq.~(\ref{2.2.3}).

In terms of the four characteristic lengths -- 
the microscopic lattice constant (unity), the critical droplet radius $R_c$, 
the average droplet distance $R_0$, and the system size $L$ -- the 
domain of validity of the KJMA approximation can be summarized as 
\begin{equation}
\label{eq:intro2}
1 \ll R_c \ll R_0 \ll L \;.
\end{equation}

Setting $R_0 \approx L$, one obtains the crossover field that limits the 
MD regime in the weak-field/small-system direction, called 
the {\it dynamic spinodal\/} (DSP):\cite{RIKV94A,TOMI92A} 
\begin{equation}
\label{eq:HDSP}
H_{\rm DSP}(T,L) \sim 
\left( \frac{1}{d+1} \frac{\Xi(T)}{\ln L}\right)^{\frac{1}{d-1}} \,.
\end{equation}
Its exceedingly slow asymptotic convergence with $L$ 
results from Eqs.~(\ref{eq:Ia}) and~(\ref{eq:R0}). However, 
relatively large systems ($L \agt 10^3$$-$10$^4$ for $d$=2) are 
required for the contribution described by Eq.~(\ref{eq:HDSP}) to 
be larger than the various correction terms (see Fig.~11 of 
Ref.~\onlinecite{JLEE94A}).

This discussion reveals three restrictions that pertain to the 
applicability of the KJMA picture to
real systems, as well as to the Ising lattice-gas model:
\begin{enumerate}
\item 
It applies only in the MD 
regime\cite{RICH94,RIKV94,RIKV94A,TOMI92A} of large systems and/or 
intermediate fields. In this regime the decay proceeds through a large 
number of droplets which nucleate independently at random times and positions 
and subsequently grow to fill the system. 
\item 
It describes nucleation and growth in a coarse-grained sense. This means 
that the results of the theory should agree with those of the Ising 
model at length scales much larger than the critical droplet radius 
$R_{c}$ and should disagree at length scales on the order of $R_{c}$ 
and shorter. 
\item 
It does not take into account interfacial effects, except insofar as they
determine the nucleation rate. When the volume fraction of 
stable phase is large, the dynamics are dominated by droplet coalescence, 
which is accelerated by the interface tension.
We therefore expect the theory to disagree with the simulation results 
in this late-time regime. 
\end{enumerate}

\subsubsection{Two-point correlations}
\label{subsec:2.2b}

The connected two-point correlation function for the metastable phase, 
$\Gamma(\vec{r},t)$, was obtained in closed form by Sekimoto under the 
assumption that the droplets are $d$-dimensional 
spheres ($\Omega_2 \! = \! \pi$, 
$\Omega_3 \! = \! 4 \pi /3$):\cite{SEKIMOTO} 
\begin{eqnarray}
\Gamma(\vec{r},t) 
& \equiv & 
\langle u(\vec{x},t)u(\vec{x} 
+ \vec{r},t) \rangle - \langle u(\vec{x},t) \rangle^{2} \nonumber \\
            & = & \left\{ \begin{array}{ll}
\langle u(t) \rangle^{2} \left\{ \exp \left[ Iv^{d}t^{d+1} \Psi_d(r/2vt)
       \right] -1 \right\},        & \mbox{$r < 2vt$} \\
        0,                         & \mbox{$r > 2vt$} 
         \end{array}
       \right. \nonumber\\
& &
\label{2.2.4}
\end{eqnarray}
where $r$=$|\vec{r} \, |$ and
\begin{mathletters}
\begin{eqnarray}
\Psi_2(y) &=& \frac{2}{3} \Bigg[ \arccos y - 2y \sqrt{1-y^{2}} \nonumber\\ 
          & & + y^{3} \ln
          \left( \frac{1+\sqrt{1-y^{2}}}{y} \right) \Bigg]  \\
\label{2.2.5a}
\Psi_3(y) &=& \frac{\pi}{3} (1-y)^3 (1+y) \,.
\label{2.2.5b}
\end{eqnarray}
\end{mathletters}
The first moment of $\Gamma(r,t)$ is defined by
\begin{equation}
\langle r(t) \rangle = \frac{\int r\; \Gamma(r,t)\; dr}
{\int \Gamma(r,t)\; dr}\, ,
\label{2.2.6}
\end{equation}
consistent with Eq.~(\ref{2.1.6}). This is a time-dependent characteristic  
length which describes the structure of the system. One would expect its 
value at $t \! = \! \tau$ to 
be proportional to $R_0$, which is confirmed by our simulations. 

\subsubsection{Relations between KJMA and Ising quantities}
\label{subsec:2.2c}

Theoretical approximate expressions for the relaxation function $\phi(t)$ and 
the correlation function $G(r,t)$ of the Ising model can be derived from the 
corresponding quantities in the KJMA theory. The main assumption is that for 
sufficiently late times, when the mean size of the droplets of stable phase is 
much larger than $R_{c}$, the mesoscopic structure 
of the Ising model resembles the KJMA picture. 
This assumption is also expected to break down at late times, when droplet 
coalescence becomes important.

The time-dependent magnetization of the Ising model, 
$m(t)$, is approximately given in
terms of the volume fraction in the KJMA theory as
\begin{equation}
m(t) \approx [m_{\rm ms} - m_{\rm s}] \, \varphi_{\rm ms}(t) + m_{\rm s} \; ,
\label{2.2.7}
\end{equation}
where $m_{\rm ms}$ and $m_{\rm s}$ are the magnetizations of the domains of 
metastable and stable phase, respectively. From the definition of the 
relaxation function $\phi(t)$, Eq.~(\ref{2.1.3}), one also has
\begin{equation}
m(t) = [m(0) - m_{\rm s}] \, \phi(t) + m_{\rm s}
\label{2.2.8}
\end{equation}
for the Ising model. Equations~(\ref{2.2.7}) and~(\ref{2.2.8}) together yield
\begin{equation}
\phi(t) \approx \frac{m_{\rm ms} - m_{\rm s}}{m(0) - m_{\rm s}} \, 
\varphi_{\rm ms}(t) 
\label{2.2.9}
\end{equation}
for the relaxation function $\phi(t)$ of the Ising model in terms of 
$\varphi_{\rm ms}(t)$ in the KJMA theory. The right-hand side of
Eq.~(\ref{2.2.9}) can be considered as a coarse-graining approximation for
$\phi(t)$, in which the local spins have been averaged over a region which 
is large compared to $R_c$, but small compared to $R_0$. 

At this level of coarse-graining, 
the correlation function of the Ising model is given in terms of the 
corresponding quantity in the KJMA theory, $\Gamma(\vec{r},t)$, as 
\begin{eqnarray}
G(\vec{r},t) 
       & \approx & 
       \left [ m_{\rm ms} - m_{\rm s} \right ]^{2} \Gamma(\vec{r},t)
        + \widehat{G}_{\rm ms}(\vec{r}) + \widehat{G}_{\rm s}(\vec{r}) 
\;.
\label{2.2.10}
\end{eqnarray}
Here $\widehat{G}_{\rm ms}(\vec{r})$ and $\widehat{G}_{\rm s}(\vec{r})$ are 
correlation functions describing local fluctuations that are nonzero only 
in the metastable and stable phase, respectively. 
(See Appendix~\ref{sec:A1}.)
These correlation functions are of very short range compared to 
$\Gamma(\vec{r},t)$, 
and where they are different from zero they are proportional to 
$\varphi_{\rm ms}$ and $(1 \! - \! \varphi_{\rm ms})$, respectively. 

Equation~(\ref{2.2.10}) enables us to obtain a KJMA approximation for the 
variance of the Ising magnetization.
The variance is obtained from the correlation function as 
\begin{equation}
\mbox{Var}[m(t)] = L^{-d} \sum_{i=1}^{L^d} G(\vec{r}_{i},t) \; .
\label{4.2.1}
\end{equation}
Combining Eqs.~(\ref{2.2.4}), (\ref{2.2.10}), and~(\ref{4.2.1}), 
we obtain\cite{RICH94} 
\begin{eqnarray}
L^d \mbox{Var}[m(t)] 
& \approx & \left [ m_{\rm ms} - m_{\rm s} \right ]^{2} 
d \Omega_d \left(2 v t \right)^d \varphi_{\rm ms}^{2}(t) \nonumber \\
& & \times \left [ \Theta_d(I v^{d} t^{d+1}) - \frac{1}{d} \right ] \nonumber\\
& & + \varphi_{\rm ms}(t) \, k_{\rm B} T \chi^{\rm ms}_{T} \nonumber\\
& &    + [1 - \varphi_{\rm ms}(t)] \, k_{\rm B} T \chi^{\rm s}_{T} \; ,
\label{4.2.2c}
\end{eqnarray}
where the function  
\begin{equation}
\Theta_d(x) 
 = \int_{0}^{1} y^{d-1} e^{x \Psi_d(y)} {d} y 
\label{4.2.2b}
\end{equation}
is obtained by numerical integration. 
Here $\chi^{\rm s}_{T}$ is the isothermal susceptibility in the 
equilibrium phase, and $\chi^{\rm ms}_{T}$ 
can be interpreted as an analogous measure of
the subcritical fluctuations in the metastable phase.

\section{Numerical Results}
\label{sec:3}

In this Section we present our simulation results for the 
two-dimensional Ising lattice-gas model and use
them to obtain the parameters in the theoretical predictions of the 
extended KJMA 
theory.\cite{KOLM37,JOHN39,AVRAMI,EVAN45,SEKIMOTO,OHTA87} 
The relaxation function, which is a one-point function, is discussed in 
Sec.~\ref{subsec:3.1}. Those quantities which also require knowledge
of two-point correlation functions are treated in 
Secs.~\ref{subsec:4.2}--\ref{subsec:4.5}. 
In the remainder of 
this paper we use units such that $J \! = \! k_{B} \! = \! 1$ 
and measure time in Monte Carlo steps per site (MCSS). 

All the results shown correspond to $T \! = \! 0.8T_{\rm c}$, 
and most of them are for $L \! = \! 256$. Only for the two weakest fields, 
$|H| \! = \! 0.15$ and~0.12, did we use $L \! = \! 1024$ to ensure a 
sufficiently large value of $L/R_0$.
Results are averaged over 100 independent realizations, 
except for $|H| \! = \! 0.12$, for which only 50 realizations were performed 
due to the long lifetime at this weak field. 

\begin{figure}
\includegraphics{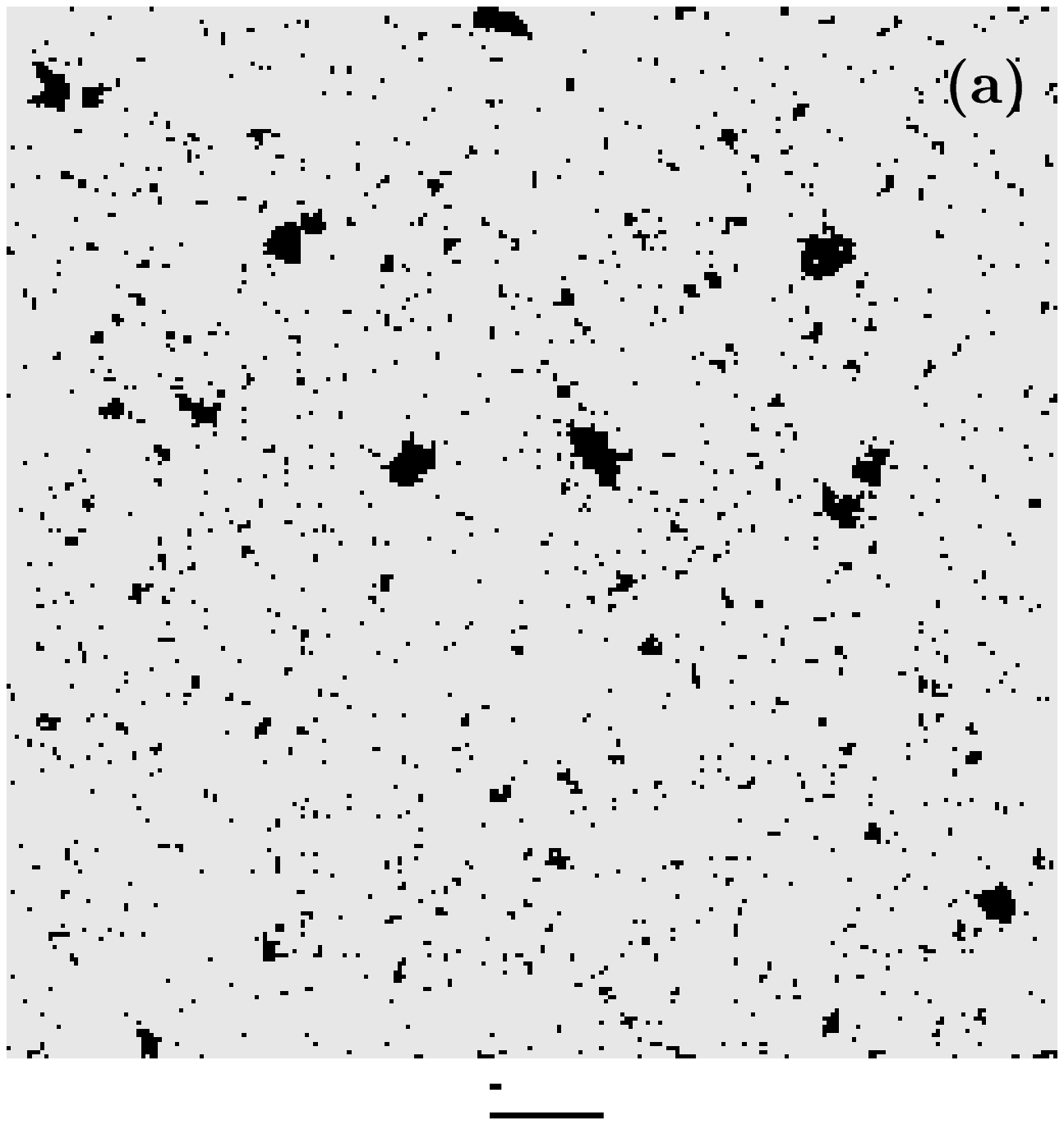}
\includegraphics{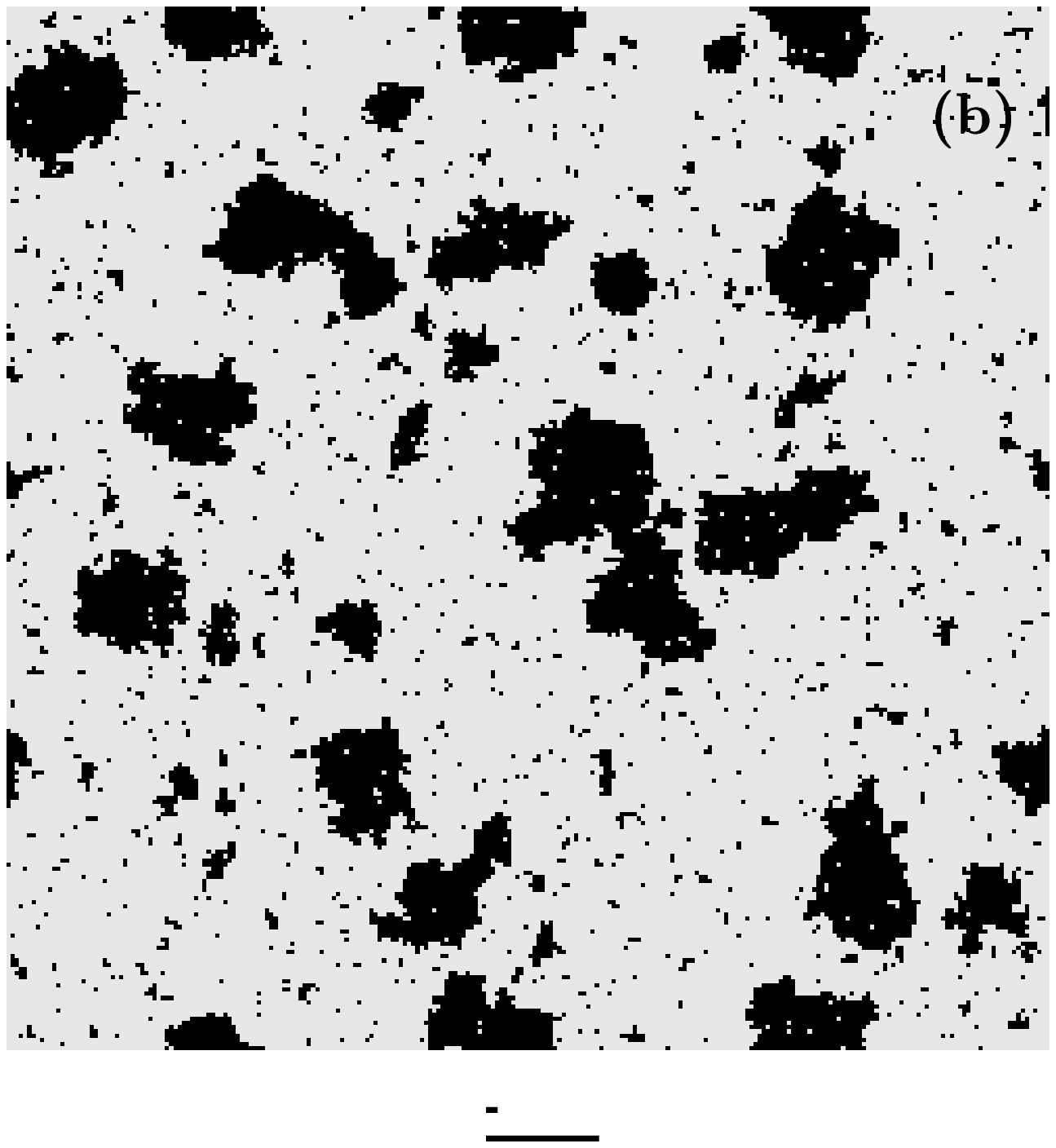}
\includegraphics{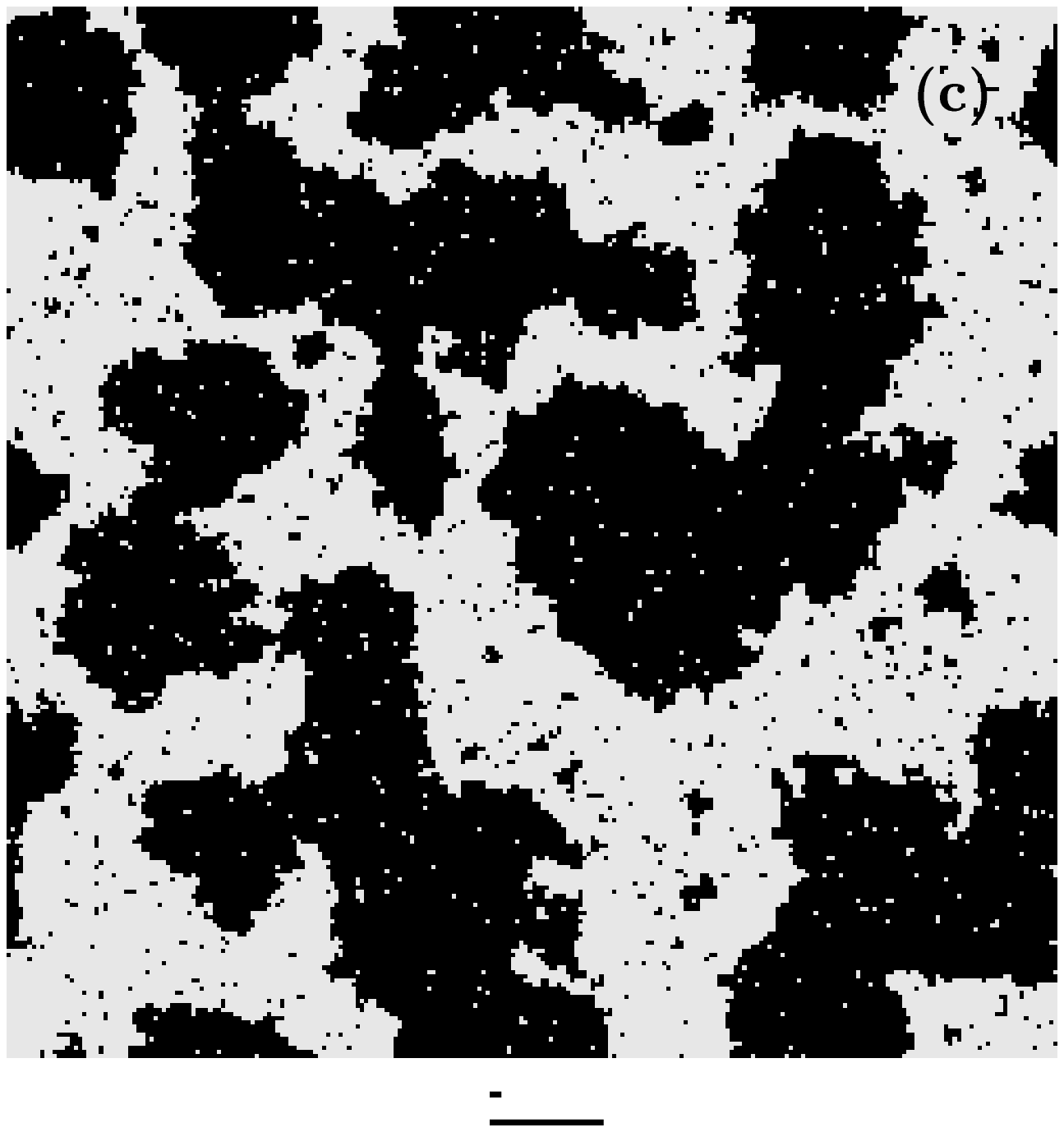}
\caption[]{
Snapshots showing the time evolution of the system configuration during a 
typical Monte Carlo (MC) 
simulation for $L$=250, $T$=0.8$T_{\rm c}$, and $|H|$=0.15, 
for which the lifetime $\tau \approx 392$~MCSS. 
Light gray represents the metastable spin direction and black the stable 
spin direction. 
While supercritical droplets in one snapshot can also be identified in the 
next, the subcritical fluctuations are uncorrelated between snapshots. 
The two characteristic lengths, $R_0 \! \approx \! 25$ and 
$R_c \! \approx \! 2.5$, are shown by the long and short bar below 
each snapshot, respectively. 
(a) $t = 80$~MCSS. 
A few supercritical droplets are seen, 
but most of the black pixels represent subcritical fluctuations. 
(b) $t = 260$~MCSS. 
Many large supercritical droplets are growing 
against the metastable background. 
(c) $t = 390$~MCSS$ \approx \tau$. 
The magnetization is close to zero, several of the 
supercritical droplets have coalesced, and the stable phase is close to 
percolating. Microscopic equilibrium fluctuations appear as light specks 
inside the stable-phase regions. 
Figure courtesy of G.~Korniss.  
}
\label{fig_snap}
\end{figure}

The evolution of the system geometry 
during the decay process is illustrated in Fig.~\ref{fig_snap} 
by a series of typical simulation snapshots for $|H|$=0.15. 

Values at 0.8$k_{\rm B}T_{\rm c} \! \approx \! 1.815\,$\cite{ONSA44} 
of quantities for the Ising model, that are needed to compare the simulations 
with the KJMA predictions are as follows. 
The surface tension 
$\sigma_0 \! \approx \! 0.745$\cite{ONSA44} 
and the zero-field magnetization 
$m_{\rm sp} \! \approx \! 0.954$.\cite{YANG52} 
The corresponding values of $\Omega_d$ and $\Xi(T)$ are 
$\Omega_2 \! \approx \! 3.153 \! \approx \! \pi$ and 
$\Xi \! \approx \! 0.506$.\cite{CCAG94A} 
This value of $\Omega_2$ gives 
the constant in the definitions of $t_0$ and $R_0$ as $A \! \approx \! 0.870$, 
and it implies that the average deviations 
of the droplets of stable phase from the circular shape assumed by the 
KJMA theory should be negligible at this temperature. 
Considering the
irregular shapes of the individual droplets in Fig.~\ref{fig_snap}, we 
find it quite remarkable that the extended KJMA theory nevertheless 
gives a very good description of the decay process, as we now proceed 
to demonstrate.

\subsection{Relaxation Function and Metastable Magnetization}
\label{subsec:3.1}

\begin{figure}
\epsfxsize 3.5in \epsfbox{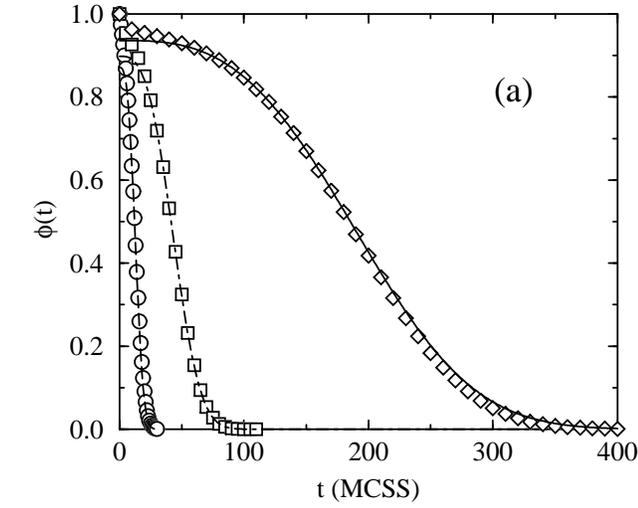}
\epsfxsize 3.5in \epsfbox{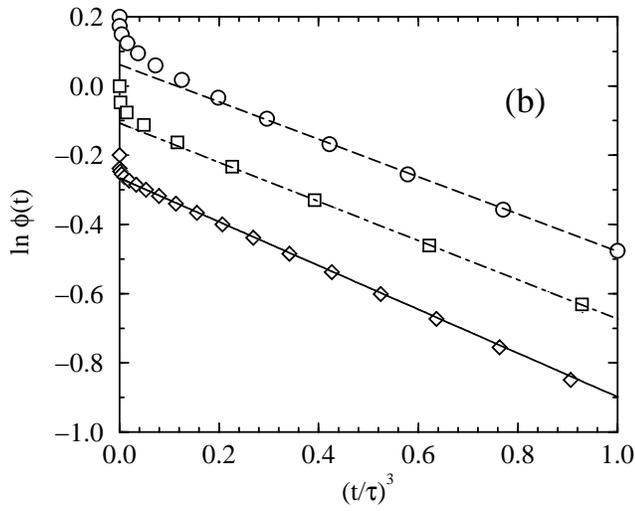}
\epsfxsize 3.5in \epsfbox{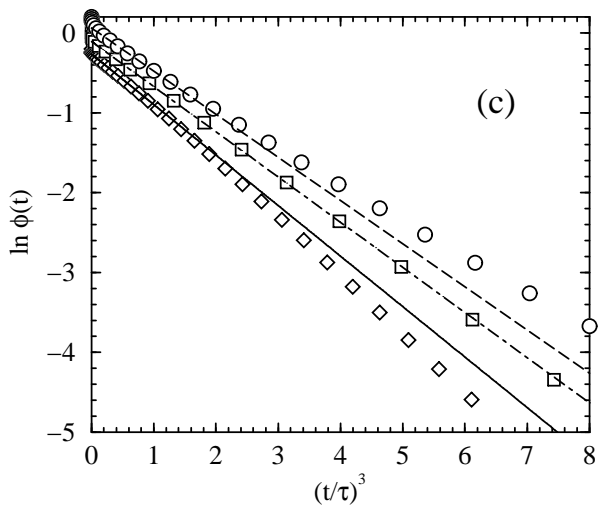}
\caption[]{
MC (points) and fitted KJMA results (lines) for the relaxation 
function of the Ising model, $\phi(t)$, for $L$=256 and $T$=0.8$T_{\rm c}$. 
Results are shown for $|H|$=0.2 ($\Diamond$ and solid curve) and 
0.4 ($\Box$ and dot-dashed curve), both in the MD regime, 
and for $|H|$=0.8 ({\protect\large $\circ$} and dashed curve), 
slightly beyond the mean-field spinodal.  
The lifetimes are $\tau \approx 186$ MCSS for $|H|$=0.2,
$\tau \approx 41$ MCSS for $|H|$=0.4, and $\tau \approx 12$ MCSS for 
$|H|$=0.8. 
(a) 
Linear scale vs $t$.
Derivatives of data such as these provide estimates for transient currents 
in ferroelectric switching\protect\cite{MITO94} 
and electrochemical potential-step experiments.\protect\cite{RIKV97,RIKV98} 
(b)
Semi-logarithmic scale vs $(t/\tau)^3$. 
{}For clarity, the data for $|H|$=0.8 and
0.2 have been displaced by $\pm 0.2$, respectively. 
(c) 
Same as (b), but plotted for $t/\tau \le 2$ to show the expected
deviations of the MC data from the KJMA approximation for later times. 
}
\label{fig_relax1}
\end{figure}

Monte Carlo (MC) and fitted KJMA results for the relaxation function 
$\phi(t)$ are shown together in Fig.~\ref{fig_relax1} for 
$|H| \! = \! 0.2$ and~0.4 (both in the MD regime), 
and~0.8 (slightly beyond the mean-field spinodal). 
The results are shown on a linear scale in Fig.~\ref{fig_relax1}(a), 
while the linear dependence of $\ln\phi(t)$ on $(t/\tau)^{3}$ predicted by 
Eq.~(\ref{2.2.3}) is illustrated in Fig.~\ref{fig_relax1}(b). 
The KJMA expression for $\phi(t)$ contains two parameters, 
$m_{\rm ms}$ and $I v^2$, which were 
determined by fitting to the MC data as described below. 

As discussed in Sec.~\ref{subsec:2.2}, the theoretical 
expression for $\phi(t)$ follows from the assumption that for times when 
the mean size of the domains of stable phase is much larger than the critical 
droplet size $R_{c}(H,T)$, the mesoscopic structure of the Ising model 
resembles the coarse-grained KJMA picture. This 
assumption leads to Eq.~(\ref{2.2.9}), from which the following 
two-parameter expression for $\ln \phi(t)$ results:
\begin{equation}
\ln \left[ \phi(t;a(|H|),b(|H|)) \right] 
\approx \ln \left[ a(|H|) \right] -b(|H|)\,t^{3} \; ,
\label{3.1.1}
\end{equation}
where $a(|H|)$ contains information about the magnetization of the
metastable phase and is given by
\begin{equation}
a(|H|) = \frac{m_{\rm ms}-m_{\rm s}}{m(0)-m_{\rm s}} \; ,
\label{3.1.2}
\end{equation}
and $b(|H|)$ is obtained from Eq.~(\ref{2.2.3}) as
\begin{equation}
b(|H|) = \frac{\Omega_2 I v^{2}}{3} \;.
\label{3.1.3}
\end{equation}

The theoretical results for 
$\phi(t)$ were obtained by performing unweighted linear least-squares fits of 
Eq.~(\ref{3.1.1}) to the MC results for $\ln \phi(t)$. 
Since coalescence effects are expected to make  
Eq.~(\ref{3.1.1}) invalid for $m \alt 0$, 
we used only data for $t \! \leq \! \tau(|H|)$ in the fits. 
We eliminated the early-time regime of rapid approach to 
``metastable equilibrium'' [the ``hooks'' most easily seen in 
Fig.~\ref{fig_relax1}(b)] by selecting the lower limit of the 
fitting interval, $t_{\rm min}(|H|)$. 
Two different criteria were used.\\
{\it (a)\/} $t_{\rm min}(|H|)$ was selected to give a joint extremum for 
$a(|H|)$ and $b(|H|)$, yielding lower bounds for $m_{\rm ms}$ 
and $Iv^2$. This minimizes the sensitivity of the estimates to the cutoff.\\ 
{\it (b)\/} $t_{\rm min}(|H|)$ was selected to give a minimum or a plateau 
in the $\chi$-square per degree of freedom in the fit.\\ 
{}For small $|H|$ 
the difference between the estimates resulting from these two criteria
is much smaller than their individual 
statistical errors, indicating that the KJMA parameters are well defined. 
For $|H| \agt 0.3$, the time scales corresponding to 
the fast relaxation towards the metastable quasi-equilibrium 
and the slow decay towards equilibrium are not well 
separated. This results in the disappearance of the extrema used to 
select $t_{\rm min}(|H|)$ and a rapid loss of precision in the definition of 
the fitting parameters with further increase of $|H|$. 

We use unweighted fits because the values of $m(t)$ 
at different $t$ are {\it not\/} independent. 
The statistical errors in the data in the intermediate-time regime 
expected to be most compatible with Eq.~(\ref{3.1.1}) are larger than 
those in the early-time ``hook'' regime (see discussion of 
Var[$m(t)$] in Subsec.~\ref{subsec:4.3} below). 
A weighted fitting procedure 
produced much inferior agreement between the fitting function and the data  
than the unweighted procedure.
As seen from Figs.~\ref{fig_relax1}(a) and~(b), 
the agreement between the MC results and the predictions of the KJMA theory 
is excellent for intermediate times. 

The progressive breakdown of the validity of the KJMA prediction for 
$\phi(t)$ at late times is illustrated in 
Fig.~\ref{fig_relax1}(c), which shows the same data as   
Fig.~\ref{fig_relax1}(b) up to $t \! = \! 2 \tau(|H|)$. 
{}For $|H| \! = \! 0.2$, the KJMA approximation agrees very well with the 
MC data at intermediate times, whereas  
for long times it decays more slowly than the 
MC results. This is 
expected since the KJMA approximation does not incorporate the 
interface-tension effects which accelerate the decay 
in the late-time regime where droplet coalescence 
becomes important. For $|H| \agt 0.5$, the KJMA approximation for 
$\phi(t)$ agrees well with the MC data at 
intermediate times, whereas for late times it decays {\em faster\/} 
than the MC results, as illustrated by the data for $|H| \! = \! 0.8$. 
This qualitative change in the 
late-time behavior of $\phi(t)$ signals the breakdown of the KJMA 
nucleation-and-growth picture as $|H|$ approaches the mean-field spinodal 
($H_{\rm MFSP} \approx 0.75$ at 0.8$T_{\rm c}$). 
In the strong-field regime beyond $H_{\rm MFSP}$ the nucleation of 
very small droplets of stable phase becomes the dominant decay mechanism, 
rather than the growth of larger domains. 
The almost perfect agreement between the Ising and KJMA results for 
$|H| \! = \! 0.4$ we believe to be the result of an accidental cancellation 
of corrections at late times. 

\begin{figure}
\epsfxsize 3.5in \epsfbox{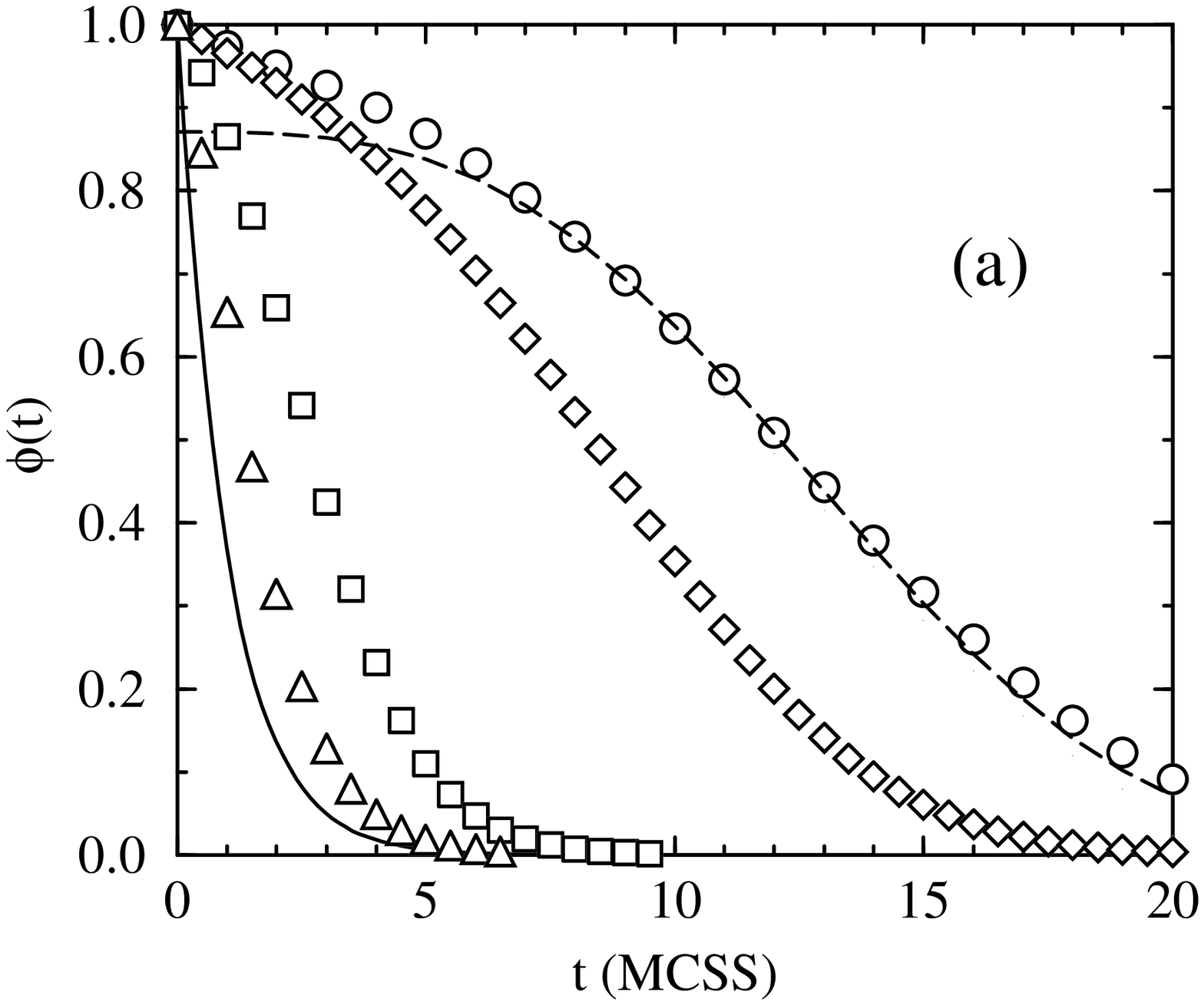}
\epsfxsize 3.5in \epsfbox{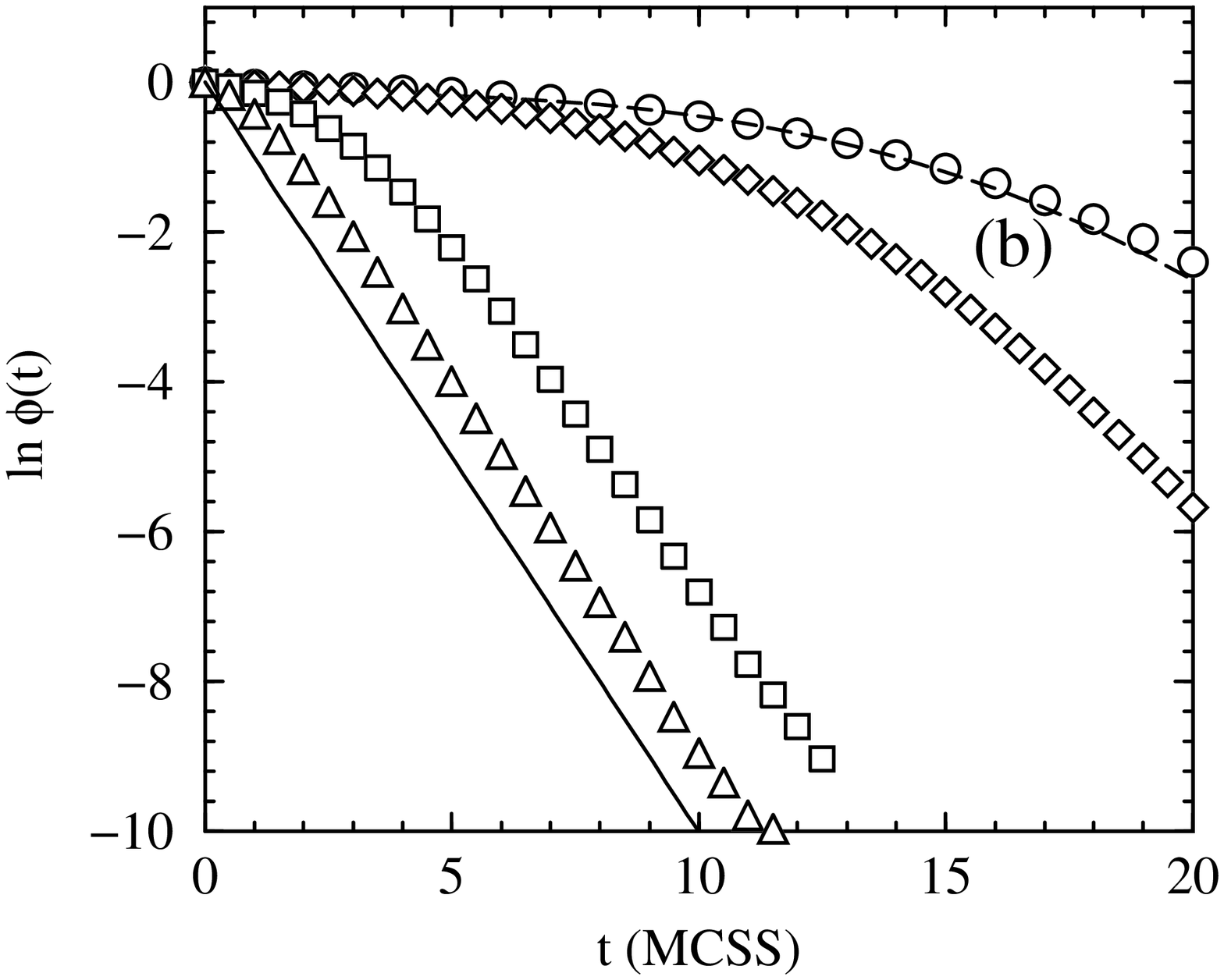}
\caption[]{
MC results for $\phi(t)$
in the strong-field regime at $|H|$=0.8 ({\protect\large $\circ$} 
and dashed curve, 
repeated from Fig.~\protect\ref{fig_relax1} for comparison), 
1.0 ($\Diamond$), 2.0 ($\Box$), and 3.0 ($\triangle$). 
The lifetimes are $\tau \approx 8.42$ MCSS for $|H|$=1.0,
$\tau \approx 2.70$ MCSS for $|H|$=2.0, and $\tau \approx 1.42$ MCSS for
$|H|$=3.0.
The solid curves represent exponential relaxation, which is the exact 
result in the limit $|H| \rightarrow \infty$. 
(a) 
Linear scale.
(b)
$\ln \phi(t)$ vs $t$, emphasizing the approach towards exponential 
decay with increasing $|H|$. 
}
\label{fig_relax2}
\end{figure}

Monte Carlo data for $\phi(t)$ in the strong-field regime, 
at $|H| \! = \! 0.8$, 1.0, 2.0, and 3.0, 
are shown in Fig.~\ref{fig_relax2}. 
As expected, the MC data are not well approximated by the 
KJMA result in this regime. 
The solid curve is the exact limit for $|H| \rightarrow \infty$; 
$\phi(t) = \exp(-t)$. The data for $|H| \! = \! 3.0$ are close to this limit.

{}For the two weakest fields, $|H| \! = \! 0.15$ and~0.12, we noticed a 
slight increase in the minimum $\chi$-square per degree of freedom in the 
fits. This may indicate that for even weaker fields 
one may need to consider the ``incubation time'' for near-critical 
clusters, which is 
discussed by Shneidman and coworkers.\cite{SHNE98,WEIN97,SHNE93X} 
Investigations for weaker fields are therefore desirable.  

\begin{figure}
\epsfxsize 3.5in \epsfbox{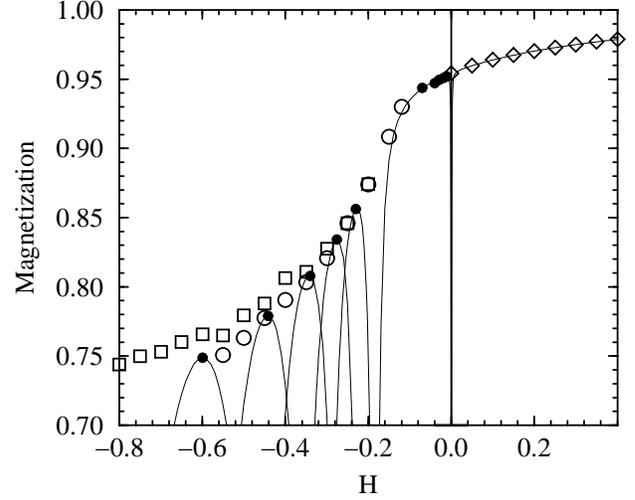}
\caption[]{
Stable magnetization, $m_{\rm s}$ ($\Diamond$), 
and metastable magnetization, 
$m_{\rm ms}$ ({\protect\large $\circ$} and $\Box$), 
shown vs $H$. 
The data points for $m_{\rm ms}$ represented as 
{\protect\large $\circ$} are 
based on the selection criterion {\it (a)\/} 
in Sec.~{\protect\ref{subsec:3.1}}, 
and those represented as $\Box$ on criterion {\it (b)\/}.
The data are for $L \! = \! 256$ except for $H \! = \! -0.15$ and~$-$0.12, 
which correspond to $L \! = \! 1024$.
The thin solid curves are transfer-matrix (TM) results. For $H \! > \! 0$ 
the curve represents the equilibrium magnetization as obtained for an 
$N \! \times \! \infty$ system with $N \! = \! 9$, which is seen to be in 
complete agreement with the MC results. 
{}For $- 0.2 \! < \! H \! < \! 0$ the curve represents the 
metastable magnetization 
corresponding to ``the first lobe'' [see 
Appendix~\protect\ref{sec:ATM}] for $N \! = \! 9$, while the solid circles 
represent the maximum of this lobe 
{}for $N \! = \! 5$, \dots,~9 from left to right. 
{}For $H \! < \! -0.2$ the curves and solid circles represent 
``the second lobe'' for $N \! = \! 5$, \dots,~9 from left to right. 
}
\label{fig_mh}
\end{figure}

Using Eq.~(\ref{3.1.2}) together with the fitting parameters $a(|H|)$, 
we obtained estimates for the 
metastable magnetization $m_{\rm ms}$ as a function of $H$. 
The equilibrium magnetizations $m_{\rm s}$ for each value 
of $H$, which are necessary inputs for this calculation, 
were obtained by standard equilibrium MC simulation.\cite{EQUILMC} 
They are shown in the right-hand part of Fig.~(\ref{fig_mh}).  

The estimates for $m_{\rm ms}$ are 
shown in the left-hand part of Fig.~(\ref{fig_mh}). The statistical errors 
for these estimates (as for most of the other estimates of nonequilibrium 
quantities presented in this paper) were calculated by dividing the set of 
independent runs into five equal batches. 
They are everywhere smaller than the symbol size and therefore not shown. 
The metastable magnetization is the quantity which is most sensitive to the 
short-time cutoff $t_{\rm min}(|H|)$ used in the fitting process. For 
$|H| \alt 0.25$, the two estimates agree to within the (small) statistical 
error. For stronger fields, the estimates differ, indicating 
that $m_{\rm ms}(|H|)$ becomes increasingly ill defined as the  
field is increased. This is the main source of uncertainty in our estimates of 
the metastable magnetization. 

The metastable magnetizations shown in Fig.~\ref{fig_mh} approach the 
curve of equilibrium magnetizations in a fashion that appears quite smooth 
and resembles an analytic 
continuation.\cite{RIKV94,CCAG93,CCAG94A,LANGER,GNW80,HARR84} 
We find this resemblance 
quite remarkable, since these estimates are obtained from 
observations of the time-dependence of the decay process, effectively 
using the theoretical KJMA result for $\phi(t)$ to subtract those regions of 
the system which have already decayed into the stable phase at any 
particular time. Somewhat fancifully, this might be 
called ``analytic continuation on the fly.''\cite{SEKIP} 

To check the KJMA estimates for $m_{\rm ms}$, we 
choose the transfer-matrix (TM) method first suggested and developed by 
Schulman and 
collaborators.\cite{NEWM77,MCCR78,PRIVMAN} 
A brief description of the method with details of its 
application to the present problem is given in Appendix~\ref{sec:ATM}. 
The transfer-matrix estimates for $m_{\rm ms}$, based on 
$N \! \times \! \infty$ Ising systems with $N \! = \! 5$, \dots,~9, 
are shown in Fig.~\ref{fig_mh} as solid black points. The agreement is 
gratifying and indicates that the metastable order parameter estimates 
extracted from the dynamic MC simulations using the KJMA theory are 
consistent with a different theoretical approach which is completely 
independent of the dynamics.

\subsection{Growth Velocity and Nucleation Rate}
\label{subsec:4.2}

\begin{figure}
\epsfxsize 3.5in \epsfbox{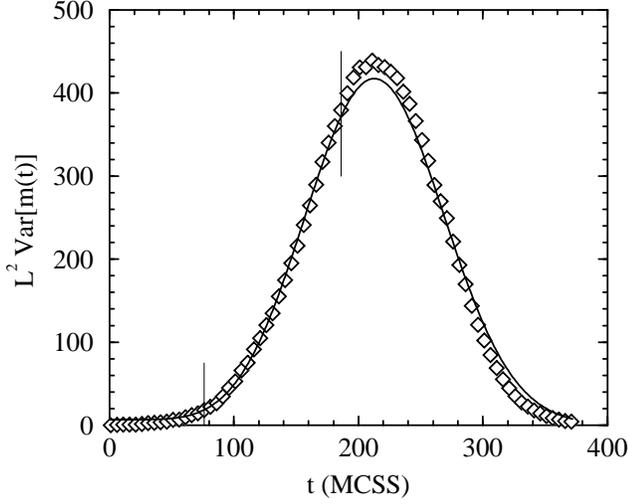}
\caption[]{
An example of $L^2$Var[$m(t)$] as obtained from a comparison of 100 
independent MC simulations ($\Diamond$) and from a least-squares fit
of the KJMA theoretical expression, Eq.~(\protect\ref{4.2.2c}), 
(solid curve). This particular result corresponds to $|H|$=0.2. 
The thin vertical lines mark the time interval $[t_{\rm min}, \tau]$, 
over which the fitting was performed. 
}
\label{fig_varm}
\end{figure}

On the basis of the relaxation function alone, one can obtain 
the combination $I v^2$ of the nucleation rate and the radial 
growth velocity from the fitting parameter $b(|H|)$. 
To obtain separate estimates for $I$ and $v$, 
one needs to compare the MC results for 
the variance of the magnetization, $\mbox{Var}[m(t)]$, with the corresponding 
KJMA prediction given in Eq.~(\ref{4.2.2c}). 
If we use the values of $m_{\rm ms}$ and $Iv^2$ obtained from the fits to 
the relaxation function, and $\chi^{\rm s}_{\rm T}$ as 
obtained from the fluctuations in the equilibrium simulations,
then $v^{2}$ and $\chi^{\rm ms}_{\rm T}$ may be determined from a 
linear fit of Eq.~(\ref{4.2.2c}) to the MC data for $\mbox{Var}[m(t)]$ 
at each value of $|H|$. 
For the same reasons discussed in the context of the fits to the relaxation
function in Sec.~\ref{subsec:3.1}, we found that an unweighted fitting 
procedure was more stable and yielded better overall agreement 
than weighted fitting. 
{}For each value of $|H|$, the values of Var[$m(t)$] were fitted over the 
same time interval as the corresponding relaxation function, and 
error bars were estimated in the same way as for $m_{\rm ms}$. 
An example of such a fit is shown in Fig.~\ref{fig_varm}. 

In Fig.~\ref{fig_vh} we show the fitted values of 
the radial growth velocity $v(|H|)$ for $|H|$ between 0.12 and 0.8. 
Since this method 
is a rather indirect way to obtain the average velocity of 
a convoluted, driven interface, one may reasonably ask whether the fitted 
$v(|H|)$ is anything more than a phenomenological parameter. 
In order to answer this question, we performed additional  
numerical and theoretical analyses as discussed in the next two paragraphs. 

In order to further test the identification 
of $v(|H|)$ with an average interface velocity,
we performed additional MC simulations of the 
time evolution of plane interfaces driven by an applied 
field.\cite{DEVI91}  
We started with $64 \! \times \! 64$ systems with all spins antiparallel 
to the applied field, except for one row of overturned spins 
along one of the lattice edges, 
and periodic boundary conditions in the direction parallel to the resulting 
interface. We then let the systems evolve in an applied field according 
to the Glauber transition probability, Eq.~(\ref{2.1.2}) 
with $T$=0.8$T_{\rm c}$ as before, except for the 
following essential modification. Nucleation in the bulk metastable 
phase was suppressed by setting equal to zero the 
transition probability of any spin parallel to all of its nearest 
neighbors.\cite{SHNEtoo} 
To distinguish them from the unconstrained growing interfaces 
discussed elsewhere in this paper, we call these interfaces ``tame.'' 
We performed 100 independent simulations, continuing each until the interface 
touched the opposite wall of the simulation box. 
The average interface position in the growth direction was calculated 
from the time-dependent magnetization as $y(t) = [m(t)+1] L_y / 2$, where 
$L_y = 64$ is the extent of the simulation lattice in the growth direction. 
Velocity estimates, obtained from linear fits to $y(t)$ 
and averaged over the 100 independent runs, are also shown in 
Fig.~\ref{fig_vh}. 
They lie very close to a straight line through the origin, of slope 
slightly less than that obtained from the fits to the KJMA theory. 
This is a reasonable result, since the 
absence of subcritical fluctuations in the ``chilled metastable phase'' 
in front of these ``tame'' interfaces should slow down their 
progress and make their average velocity a lower 
bound for the velocity of an interface growing into a metastable phase 
with a thermal distribution of subcritical fluctuations. 

The interface of a growing Ising cluster is in the 
dynamic universality class of the Kardar-Parisi-Zhang\cite{KARD86} (KPZ) 
model.\cite{DEVI91,GROS91} 
The growth velocity of a planar interface is therefore 
expected to be linear in $H$ for weak fields, as is also predicted for 
large droplets by the Lifshitz-Allen-Cahn 
theory.\cite{GUNT83B,LIFS62,ALLE79} However, 
neither theory explicitly gives the proportionality constant, which  
should depend on the average orientation of the interface, as well as 
on the specific microscopic dynamic. 
Recently, Rikvold and Kolesik\cite{RIKV99} have developed an approximate
theory for the 
growth velocity of ``tame'' Ising interfaces, based on the solid-on-solid 
(SOS) approximation for the equilibrium interface 
structure.\cite{BURT51,AVRO82}  
This theory gives rise to the theoretical 
curves shown in Fig.~\ref{fig_vh}. The corresponding 
analytic expressions are given in Appendix~\ref{sec:A2}. 
The agreement with the MC results for the ``tame'' interfaces is excellent. 

The theoretical and numerical evidence presented here 
strongly supports the assertion that $v(|H|)$ obtained by 
fitting KJMA predictions to MC simulations 
is indeed a reasonable estimate of the average propagation  
velocity for the convoluted interfaces which separate the regions of stable 
and metastable phase. The estimate appears satisfactory, even though 
the nucleation rate in the metastable phase 
is too high to measure the growth velocity by more direct methods. 

Except at very early times, 
the magnitude of the term proportional to 
$\chi_T^{\rm ms}$ 
in Eq.~(\ref{4.2.2c}) is much smaller than the first term. 
As a result, the fitted values of $v(|H|)$ are quite insensitive to 
$\chi_T^{\rm ms}$, which shows large error bars and large fluctuations 
with respect to $|H|$. Accurate determination of 
$\chi_T^{\rm ms}$ evidently would require much larger data sets than 
used in this study. 

With the separate estimates for $b(|H|)$ and $v(|H|)$, we can easily
calculate the nucleation rate $I(T,|H|)$ from 
Eq.~(\ref{3.1.3}). The results are shown in Fig.~\ref{fig_ih}. 
Adjusting the unknown coefficient $B(T)$ in the exact asymptotic relation 
Eq.~(\ref{eq:I}), such that the theoretical line goes through the data 
point at $|H|$=0.15, we find good overall agreement. The curvature of the 
theoretical result is due to the prefactor exponent $K$=3. 

\subsection{Correlation Functions}
\label{subsec:4.3}

In Sec.~\ref{subsec:4.2} we used Var[$m(t)$], which is 
proportional to the spatial integral of 
the correlation function $G(\vec{r}_i,t)$, to determine the radial growth 
velocity $v(|H|)$. We now proceed to obtain the circularly averaged 
KJMA correlation function $G(r,t)$ from 
Eqs.~(\ref{2.2.4}) and~(\ref{2.2.10}), using the MC estimates for $a(|H|)$, 
$b(|H|)$, and $v(|H|)$. Since the in-phase correlation functions, 
$\widehat{G}_{\rm ms}(r)$ and $\widehat{G}_{\rm s}(r)$, 
are of very short range, we here set them equal to zero for nonzero $r$. 
The results for $|H|$=0.2, 0.4 and 0.8 at $t$=$\tau(|H|)$ are shown in  
Fig.~\ref{fig_gr1}, together with the corresponding MC results. 
The agreement is quite good, except for $r \! \approx \! 0$. 
This small discrepancy arises from the coarse-grained nature of the KJMA 
theory and is consistent with the very short range of the in-phase 
correlations. By comparing the theoretical and MC correlation functions in 
Fig.~\ref{fig_gr1}, we infer that the ranges of the in-phase correlation 
functions are on the order of one lattice constant. The difference between 
the theoretical and simulated correlation functions at $r$=0 gives 
an estimate of $\widehat{G}_{\rm ms}(0) + \widehat{G}_{\rm s}(0)$. 
Monte Carlo results for $G \left( r,\tau(|H|) \right)$ 
in the strong-field regime are shown 
in Fig.~\ref{fig_gr2} for comparison. Note the very short range of 
the correlations in this regime. 

The time evolution of the correlations and the breakdown of the 
agreement between the KJMA approximation and the MC data for late times and 
for increasing fields are well illustrated by the time-dependent 
characteristic length $\langle r(t) \rangle$, defined in 
Eqs.~(\ref{2.1.6}) and~(\ref{2.2.6}) for MC and KJMA, respectively. 
These quantities are shown together in Fig.~\ref{fig_rt} for 
$|H|$=0.2, 0.4 and~0.8. For early and intermediate times, the 
MC and theoretical results for $\langle r(t) \rangle$ increase approximately 
linearly with $t$ until they reach a maximum at a time somewhat beyond 
$\tau$. For late times, the MC and 
theoretical results differ considerably.
This is easily understood, since 
the long-time dynamical behavior of the Ising model is dominated by 
interface tension effects which are not included in the KJMA theory. 
The characteristic lengths obtained from the 
MC simulations are shorter than the KJMA estimates by varying amounts, 
which are less than about 0.5 for $t < \tau(|H|)$. 
We believe this reflects the short-range, in-phase correlations. 
These are ignored in the KJMA estimates, whereas they are present in the 
MC correlation functions, weighting the latter slightly towards smaller $r$. 

In Fig.~\ref{fig_lh} we show the $H$ dependence of
the most important lengths that characterize the system during the decay. 
As expected, $R_0$ and 
the maximum value of $\langle r(t) \rangle$, $r_{\rm max}(T,|H|)$, 
are proportional over the whole range of fields studied. 
The diameter of 
a critical droplet, $2 R_{c}(T,H) \approx \sigma_{0}/[m_{\rm s} |H|]$, 
is everywhere smaller than these mesoscopic characteristic lengths. 

The expression for $\Gamma(r,t)$, Eq.~(\ref{2.2.4}), can be
recast in the two-parameter scaling form, 
\begin{equation}
\Gamma(r,t)/\Gamma(0,t) = \left \{ 
\begin{array}{ll}
\left \{ \exp \left [({r_0}/R_{0})^{3} 
\Psi_2(r/2{r_0}) \right ] -1 \right \}, & \mbox{$r < 2 {r_0}$} \\
        0, & \mbox{$r > 2 {r_0}$} \end{array} \right. 
\label{4.4.1}
\end{equation}
where ${r_0}(t)$=$vt$ is proportional the average 
radius of the growing domains of stable phase.\cite{RAMO95}
This two-parameter scaling behavior is illustrated in Fig.~\ref{fig_grs}, 
which shows MC and KJMA correlation functions vs $r/2 {r_0}$ for two 
different sets of $|H|$ and $t$ in the MD regime, 
chosen such that they give the same value of ${r_0}/R_0$. 
The functions are normalized such that for both sets of parameters, 
the KJMA correlation function equals unity at $r$=0. 
The dependence on ${r_0}/R_0$ in Eq.~(\ref{4.4.1}) may explain
the breakdown with increasing volume fraction of the one-parameter
scaling in terms of $r/ {r_0}$,
recently used by Huang {\it et al.\/}\cite{HUAN98} 
for the experimentally obtained domain correlation
function of polymer films undergoing phase transformation. 

\subsection{Structure Factors}
\label{subsec:4.5}

The good agreement between the MC and KJMA results for the 
correlation functions in the MD regime should be accompanied by 
similar agreement for the structure factors. 
This is confirmed by Fig.~\ref{fig_sq1}, 
which shows the MC and KJMA results for the circularly averaged structure 
factor $S\left(q,\tau(|H|)\right)$ for $|H|$=0.2, 0.4, and 0.8. 

The extent of the agreement for small $q$ is best seen in 
Fig.~\ref{fig_sq1}(a), which shows the data on a linear scale. 
This is not surprising, since it is exactly at these mesoscopic length scales 
that the KJMA theory is expected to describe the spatial structure. 

The behavior for large $q$ is best seen in the 
log-log plots in Fig.~\ref{fig_sq1}(b). 
The KJMA correlation functions are linear for small $r$ and therefore 
agree with Porod's Law,\cite{GUIN55} 
which states that the structure factor 
for a two-phase system with interfaces of negligible thickness should 
behave as $S(q) \sim q^{-(d+1)}$ for large $q$. 
The small oscillations superimposed on the $q^{-3}$ tails 
are due to the sharp cutoff in $\Gamma$ at $r$=$2{r_0}$. 
{}For the MC data, 
the thermal fluctuations and the lattice cutoff at the Brillouin zone 
boundary causes marked deviations from Porod's law. 
Considerably weaker fields and consequently larger values of $r_{\rm max}$ 
would be necessary to obtain the separation of length scales necessary 
to observe Porod's law in a substantial range of $q$ for the MC data.

\section{Discussion}
\label{sec:4}

This paper reports a detailed theoretical and simulational study of the 
transient spatial structures that evolve during phase transformation 
driven by a difference in free-energy density between a metastable and a 
stable phase. This process is 
a prototype of metastable decay in a wide range of physical and chemical 
systems, which are commonly analyzed in terms of the KJMA theory or 
one of its many extensions and generalizations. 
The model system used in the numerical part of our study is a 
two-dimensional, kinetic Ising lattice-gas model. 
This is one of the first detailed attempts to verify the 
KJMA theory and identify the limitations to its validity, using a 
model system in which the elementary kinetic processes act on time and 
length scales much smaller than those characteristic of the 
mesoscopic stable-phase 
droplets. Since the model contains no impurities or free 
surfaces, the decay occurs via homogeneous, progressive nucleation and 
subsequent growth of droplets of the stable phase. While homogeneous 
nucleation is less common in nature than heterogeneous nucleation at 
impurities and surfaces, the limitations to the KJMA picture that we 
identify, should be valid also for these more complicated situations. 

Our numerical results confirm that the KJMA theory for the volume fractions 
of stable and metastable phase (i.e., one-point functions), together 
with Sekimoto's extensions that provide two-point correlation functions, 
give a remarkably accurate description of the decay process for a wide 
range of system parameters. This regime extends surprisingly far towards 
strong fields (large supersaturations) and the correspondingly small 
nucleation barriers. The conditions for the theory's validity
are essentially as follows.
\begin{itemize}
\item
The critical droplets of stable phase must be larger than the lattice 
constant, while at the same time much smaller than the system itself. 
\item
The system must be sufficiently large that the total number of 
droplets is large. 
\item
Due to the effects of droplet coalescence, 
the theory breaks down for late times, when the remaining metastable 
volume fraction becomes less than one half. 
\end{itemize}
When these conditions are satisfied, there is a large separation between 
the microscopic time scale and the mesoscopic time scale characteristic of 
the phase separation. Under these circumstances, 
we find the extended KJMA theory 
to be sufficiently accurate that it enables us to measure nonequilibrium 
thermodynamic quantities, including the order parameter in the metastable 
phase, the droplet nucleation rate, and the average propagation velocity 
of the convoluted interface between the two phases. We are able to verify 
the measured values by independent theoretical arguments. 
Due to the relatively 
short lifetime of the metastable phase, these quantities are not easy 
to measure directly by more traditional methods, and the methods developed 
here may therefore be useful for experimental measurements on systems 
undergoing phase transformation as well. 

While we have demonstrated excellent agreement between the extended 
KJMA theory and the kinetic Ising model in two dimensions and for moderately 
strong fields, we believe it would be useful to perform similar studies 
for weaker fields and in higher dimensions. A study for $d \! = \! 3$ is 
in progress.\cite{TOWN98}

\section{Acknowledgments}
\label{sec:5}

We acknowledge useful discussions with 
K.~Sekimoto, 
M.~Kolesik, 
V.~A.\ Shneidman, 
S.~W.\ Sides, 
H.~Tomita, 
and 
H.~L.\ Richards 
and critical reading of the manuscript by G.~Brown, G.~Kolesik, and 
S.~J.\ Mitchell. 
This work was supported in part through the National
Science Foundation Grants No.\ DMR-9315969, DMR-9634873, and DMR-9871455, 
and by Florida State
University through the Center for Materials Research and Technology 
and the Supercomputer Computations Research Institute
(Department of Energy Contract No.\ FC05-85ER25000). 
P.~A.~R.\ enjoyed hospitality and support at Kyoto University 
and at the Colorado Center for Chaos and Complexity. 
Supercomputer time at the National Energy Research Supercomputer Center 
was made available by the Department of Energy. 

\appendix
\section{Coarse-Grained Approximation for the Correlation Functions} 
\label{sec:A1}

The coarse-grained approximations for $G(\vec{r},t)$ and $L^d$Var$[m(t)]$, 
Eqs.~(\ref{2.2.10}) and~(\ref{4.2.2c}), are obtained as follows 
(with the time variable suppressed for simplicity 
of notation). We write the local spin variables as 
\begin{equation}
s_i = [m_{\rm ms} + \delta_{\rm ms}(\vec{r}_i)] u(\vec{r}_i) 
+ [m_{\rm s} + \delta_{\rm s}(\vec{r}_i) ] (1 - u(\vec{r}_i)) \;,
\label{eq:a1}
\end{equation}
where $\delta_{\rm ms}(\vec{r}_i)$ 
and $\delta_{\rm s}(\vec{r}_i)$, which both average 
to zero, are local fluctuations in the metastable and the stable 
phase regions, respectively. Assuming the local fluctuations in the two phases 
are mutually uncorrelated and uncorrelated with the phase field $u$, 
Eq.~(\ref{2.1.5a}) then gives Eq.~(\ref{2.2.10}) with 
\begin{equation}
\widehat{G}_{\rm ms}(\vec{r}) 
= \langle \delta_{\rm ms}(\vec{x}) \delta_{\rm ms}(\vec{x}+\vec{r}) \rangle
\langle u(\vec{x}) u(\vec{x}+\vec{r}) \rangle 
\label{eq:a2}
\end{equation}
and analogously for $\widehat{G}_{\rm s}(\vec{r})$. Further assuming that 
the correlation lengths for the local fluctuations are much shorter 
than for the phase field, 
\begin{eqnarray}
\widehat{G}_{\rm ms}(\vec{r}) 
&\approx& 
\langle \delta_{\rm ms}(\vec{x}) \delta_{\rm ms}(\vec{x}+\vec{r}) \rangle 
\langle u^2(\vec{x}) \rangle 
\nonumber\\
& = & 
\langle \delta_{\rm ms}(\vec{x}) \delta_{\rm ms}(\vec{x}+\vec{r}) \rangle 
\langle u(\vec{x}) \rangle 
\nonumber\\
& = & 
\langle \delta_{\rm ms}(\vec{x}) \delta_{\rm ms}(\vec{x}+\vec{r}) \rangle 
\varphi_{\rm ms}  
\label{eq:a3}
\end{eqnarray}
for all $\vec{r}$ such that 
$\langle \delta_{\rm ms}(\vec{x}) \delta_{\rm ms}(\vec{x}+\vec{r}) \rangle$
is nonzero. As a result, the spatial sum over 
$\widehat{G}_{\rm ms}(\vec{r})$ can be considered an
``isothermal susceptibility'' for the metastable phase, weighted by the 
metastable volume fraction: 
\begin{equation}
\sum_{i=1}^{L^d} \widehat{G}_{\rm ms}(\vec{r}_i) 
= \varphi_{\rm ms} k_{\rm B} T \chi_T^{\rm ms} \;. 
\label{eq:a4}
\end{equation}
The same reasoning is applied to the fluctuations in the equilibrium phase, 
yielding Eq.~(\ref{4.2.2c}).

\section{Transfer-Matrix Estimates of the Metastable Magnetization} 
\label{sec:ATM}

Here we summarize the transfer-matrix (TM) calculation used to 
check the consistency of the KJMA estimates for $m_{\rm ms}(H)$ in 
Sec.~\ref{subsec:3.1}.

The field-theoretical result for the nucleation rate used in this 
work, Eq.~(\ref{eq:I}), is proportional 
to the imaginary part of an analytic continuation of the equilibrium free 
energy into the metastable 
phase.\cite{LANGER,GNW80} 
Schulman and 
collaborators\cite{NEWM77,MCCR78,PRIVMAN} 
suggested that this analytic continuation 
could be found from some of the subdominant 
eigenvalues and eigenvectors of the TM commonly used to 
calculate the equilibrium partition function and correlation lengths 
of Ising and lattice-gas systems,\cite{DOMB60} and they successfully 
obtained the metastable order parameter for the Ising ferromagnet at 
$T \! \approx \! 0.4T_{\rm c}$.\cite{MCCR78}  
The method has later been extended to obtain the full, complex-valued 
free energy.\cite{RIKV94,CCAG93,CCAG94A} 

The dominant eigenvalue $\lambda_0$ 
of the TM for an $N \! \times \! \infty$ Ising 
system is positive and nondegenerate by the Perron-Frobenius 
theorem.\cite{DOMB60} It is related to the free energy per site as 
$F(T,H) = - (T/N) \ln \lambda_0$, and the equilibrium magnetization is 
obtained 
as $m_{\rm eq}(T,H) = - d F(T,H)/dH = \langle 0 | {\cal M} | 0 \rangle$, 
where $\cal M$ is the magnetization operator and the bra and ket are the 
left and right eigenvectors corresponding to $\lambda_0$. 
When the subdominant eigenvalues $\lambda_\alpha$ are plotted in the same 
logarithmic way used to obtain the equilibrium free energy from $\lambda_0$, 
one obtains a plot like the one shown in Fig.~\ref{fig_TMlambda}. 
With each branch in the figure one can associate a magnetization 
$m_\alpha(T,H) =  d \left[(T/N) \ln | \lambda_\alpha | \right]/dH 
= \langle \alpha | {\cal M} | \alpha \rangle$. 
The metastable branch corresponds to the union of the lowest-lying 
eigenvalue branches in the figure, which have a magnetization whose sign is 
opposite that of the equilibrium magnetization. 
It is marked with thick curve segments in Fig.~\ref{fig_TMlambda}. 
At specific values of $H$ the eigenvalues in the composite metastable branch 
undergo avoided crossings with other branches, at 
which their eigenvectors and magnetizations vary rapidly. 
The field corresponding to the 
$n$th crossing ($n \! = \! 0$ corresponds to $H_0 \! = \! 0$) 
depends on $n$ and 
the strip width $N$ approximately as $H_n \approx 2J/(N-n)$, 
with best agreement for low $T$ and small $n$.\cite{CCAG94A} 
The magnetization along the 
branch between the $(n \! - \! 1)$th and $n$th crossing is what we 
refer to as the ``$n$th lobe'' in the caption of Fig.~\ref{fig_mh}. 
The point on each lobe, where the magnetization calculated from 
that lobe has its extremum, is marked as a solid black circle 
in Fig.~\ref{fig_TMlambda}. These 
points correspond to the points similarly marked in 
Figs.~\ref{fig_mh} and~\ref{fig_lh}. 

{}For this work we numerically diagonalized the TM for $N \! = \! 5,\dots,9$,
using the subroutine {\tt jacobi}\cite{NUMRECF} in double precision on 
a DEC-{\it alpha\/} workstation and a Cray J90 supercomputer. 
As seen in Fig.~\ref{fig_mh}, we found excellent agreement between the 
KJMA results and the TM estimates from the first lobe for very weak 
fields and from the second lobe for stronger fields. 
The values of $m_{\rm ms}(H)$ extracted from the second lobe for the different 
values of $N$ agree with the KJMA estimates to within the uncertainty in the 
latter. Estimates based on the third and higher lobes give less satisfactory 
agreement.

\section{Approximate Expression for the Interface Velocity} 
\label{sec:A2}

\begin{figure}
\centerline{
\epsfxsize 3.5in \epsfbox{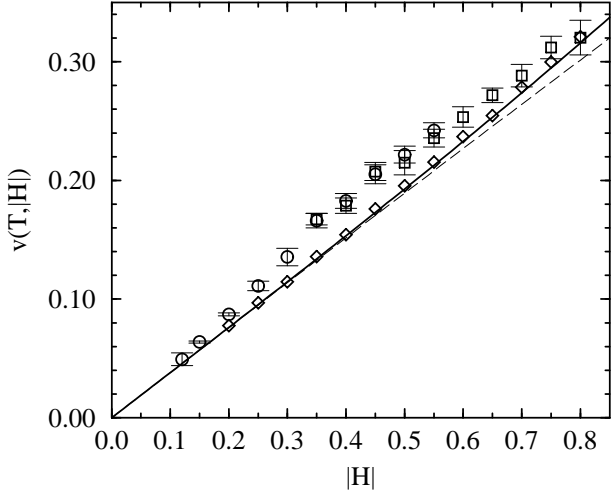}
}
\caption[]{
The average radial growth velocity of the domains of stable phase, 
$v(T,|H|)$, vs $|H|$. The circles and squares 
(corresponding to the same selection criteria as in Fig.~\protect\ref{fig_mh}) 
are the velocities
obtained by fitting Eq.~(\protect\ref{4.2.2c}) to the MC results for
$\mbox{Var}[m(t)]$. The diamonds are the velocities obtained in MC
simulations of a growing plane interface in which nucleation inside
the single-phase domains is suppressed -- a ``tame'' interface. 
The thin dashed curve and the solid curve represent the analytical 
linear-response and nonlinear-response approximations for the 
``tame'' interface 
velocities,\protect\cite{RIKV99} Eq.~(\protect\ref{eq:vSOS}). 
}
\label{fig_vh}
\end{figure}

\begin{figure}
\epsfxsize 3.5in \epsfbox{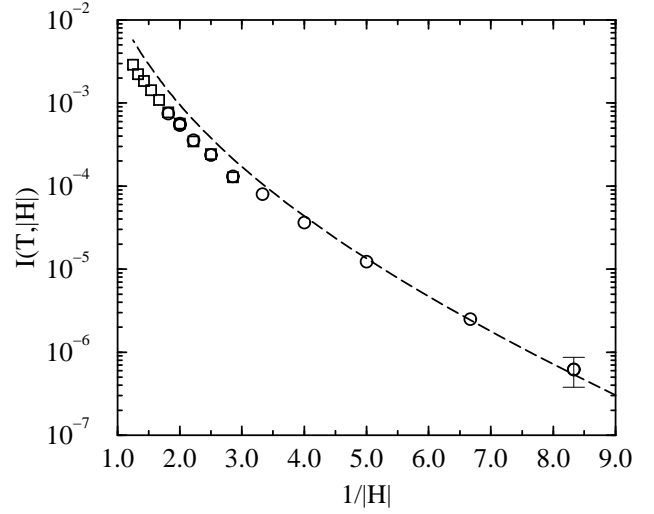}
\caption[]{
The nucleation rate $I(T,|H|)$, computed using Eq.~(\protect\ref{3.1.3}) 
with values of $b(|H|)$ and $v(|H|)$ obtained 
from fits to $\phi(t)$ and Var$[m(t)]$, respectively. 
The dashed curve is a one-parameter fit to the exact asymptotic result, 
Eq.~({\protect\ref{eq:I}}). 
}
\label{fig_ih}
\end{figure}

\begin{figure}
\epsfxsize 3.5in \epsfbox{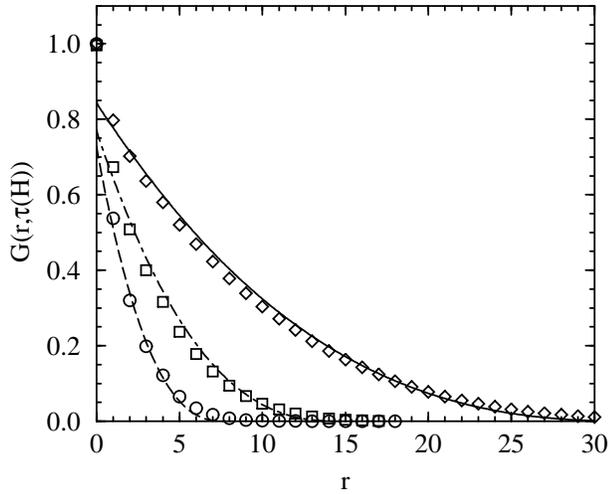}
\caption[]{
MC and KJMA results for the correlation function $G(r,\tau(|H|))$ for 
$|H|$=0.2 ($\Diamond$, solid curve), $|H|$=0.4 ($\Box$, dot-dashed curve), 
and $|H|=0.8$ ({\protect \large $\circ$}, dashed curve).
}
\label{fig_gr1}
\end{figure}

\begin{figure}
\epsfxsize 3.5in \epsfbox{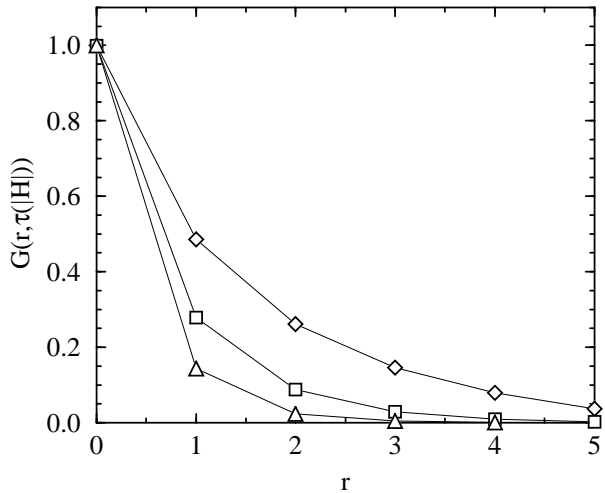}
\caption[]{
MC results for the correlation function 
$G(r,\tau(|H|))$ in the strong-field regime for $|H|$=1.0 ($\Diamond$), 
$|H|$=2.0 ($\Box$), and $|H|$=3.0 ({$\triangle$}).
The lines connecting the data points are merely guides to the eye. 
Note the very short ranges of these correlation functions. 
}
\label{fig_gr2}
\end{figure}

\begin{figure}
\epsfxsize 3.5in \epsfbox{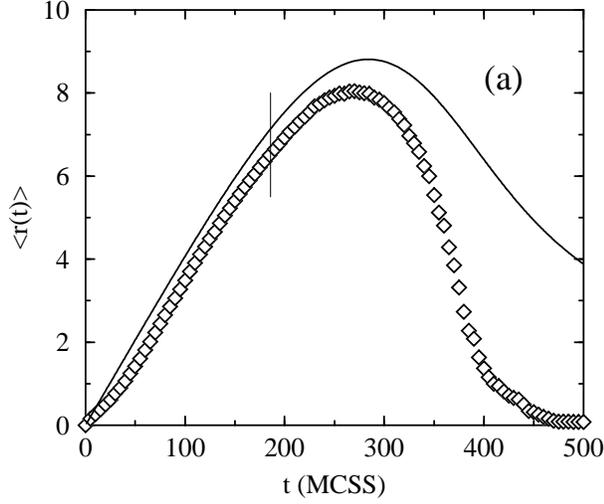}
\epsfxsize 3.5in \epsfbox{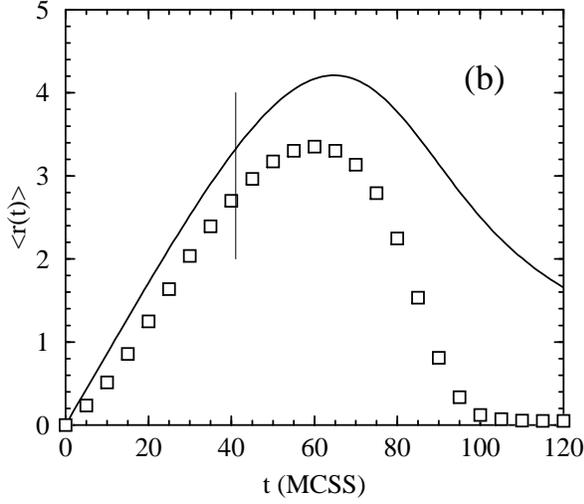}
\epsfxsize 3.5in \epsfbox{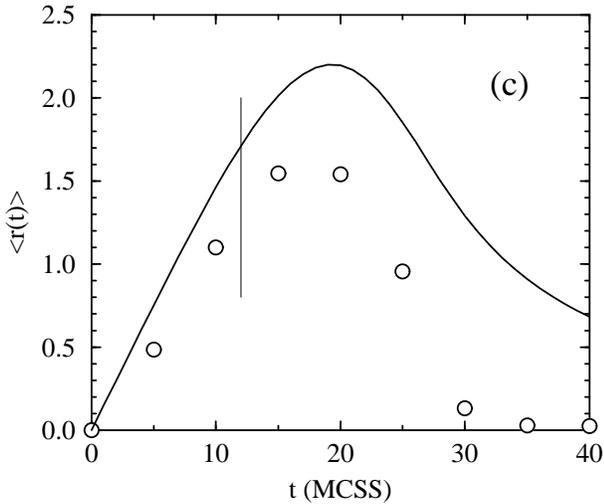}
\caption[]{
MC and KJMA results for the 
first moment of $G(r,t)$, $\langle r(t) \rangle$. 
Note the reasonable agreement at early and intermediate times. The rapid 
decrease of the MC characteristic length at late times 
reflects the acceleration of the decay of the metastable phase due to 
interface-tension effects during the droplet-coalescence regime. 
In each panel the thin vertical line marks $\tau(|H|)$. 
(a) $|H|$=0.2. 
(b) $|H|$=0.4. 
(c) $|H|$=0.8.
}
\label{fig_rt}
\end{figure}

\begin{figure}
\epsfxsize 3.5in \epsfbox{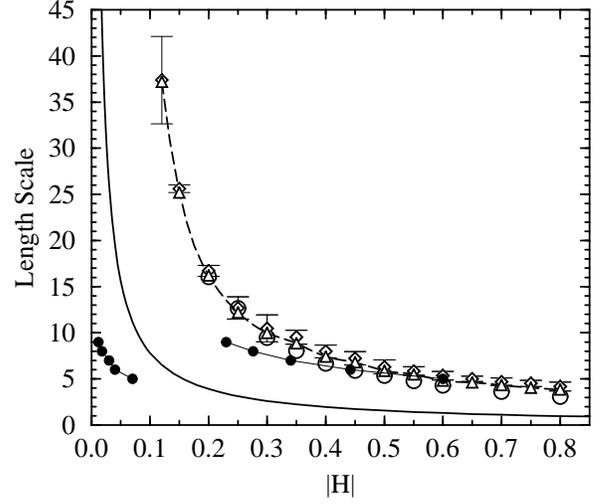}
\caption[]{
Field dependence of the characteristic lengths: 
the critical droplet diameter $2 R_{c}$ (solid curve), the mean 
separation between droplets of stable phase 
$R_{0}$ ($\Diamond$, $\bigtriangleup$, and dashed curve) 
and twice the maximum of $\langle r(t) \rangle$, 
$2r_{\rm max}$ ({\protect \large $\circ$}). 
Two sets of estimates for $R_0$ are shown: A$v/(Iv^2)^{1/3}$ 
($\Diamond$ with error bars) 
and $v \tau$ ($\bigtriangleup$ and dashed curve). 
The lengths that characterize the mesoscopic structure, 
$R_0$ and $r_{\rm max}$, 
remain proportional over the whole range of fields studied. 
The two chains of solid circles relate 
the strip widths $N$ of the transfer-matrices 
used to calculate $m_{\rm ms}$ (Sec.~\ref{subsec:3.1} 
and Appendix~\protect\ref{sec:ATM})
to the values of 
$H$ for which $m_{\rm ms}$ so calculated has a maximum. 
The data points correspond to those shown by the same symbols in 
Fig.~\ref{fig_mh}. 
The chain between $H \! = \! 0$ and~0.1 represents the first lobe, and the 
chain between 
$H \! = \! 0.2$ and~0.6 represents the second lobe; in both cases 
$N \! = \! 9$, $...$, 5 from left to right. 
Comparison of $N$ to the characteristic lengths indicate that the first 
lobe samples only subcritical fluctuations, while the second lobe 
also samples rare supercritical fluctuations in the constrained ensemble 
represented by the TM eigenspaces. 
}
\label{fig_lh}
\end{figure}

\begin{figure}
\epsfxsize 3.5in \epsfbox{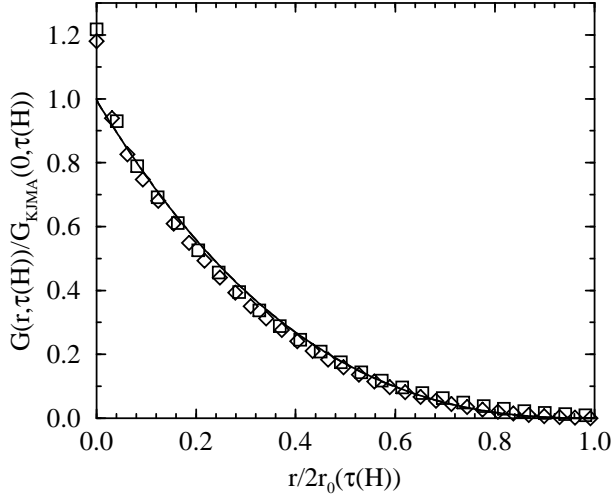}
\caption[]{
Two-parameter scaling plot of the normalized correlation function 
$G \left( r,\tau(|H|) \right) / G_{\rm KJMA} \left( 0,\tau(|H|) \right)$. 
The data points are MC results and the solid curve 
is the scaling form of the KJMA correlation function $\Gamma(r,t)$, 
Eq.~({\protect\ref{4.4.1}}). 
The MC results are shown for $|H|$=0.2 and $t$=185 
MCSS ($\Diamond$), and for $|H|$=0.25 and $t$=110 MCSS ($\Box$). 
These fields and times are chosen such that 
${r_0}/R_{0}$$\approx$0.97 
in both cases. 
}
\label{fig_grs}
\end{figure}

\begin{figure}
\epsfxsize 3.5in \epsfbox{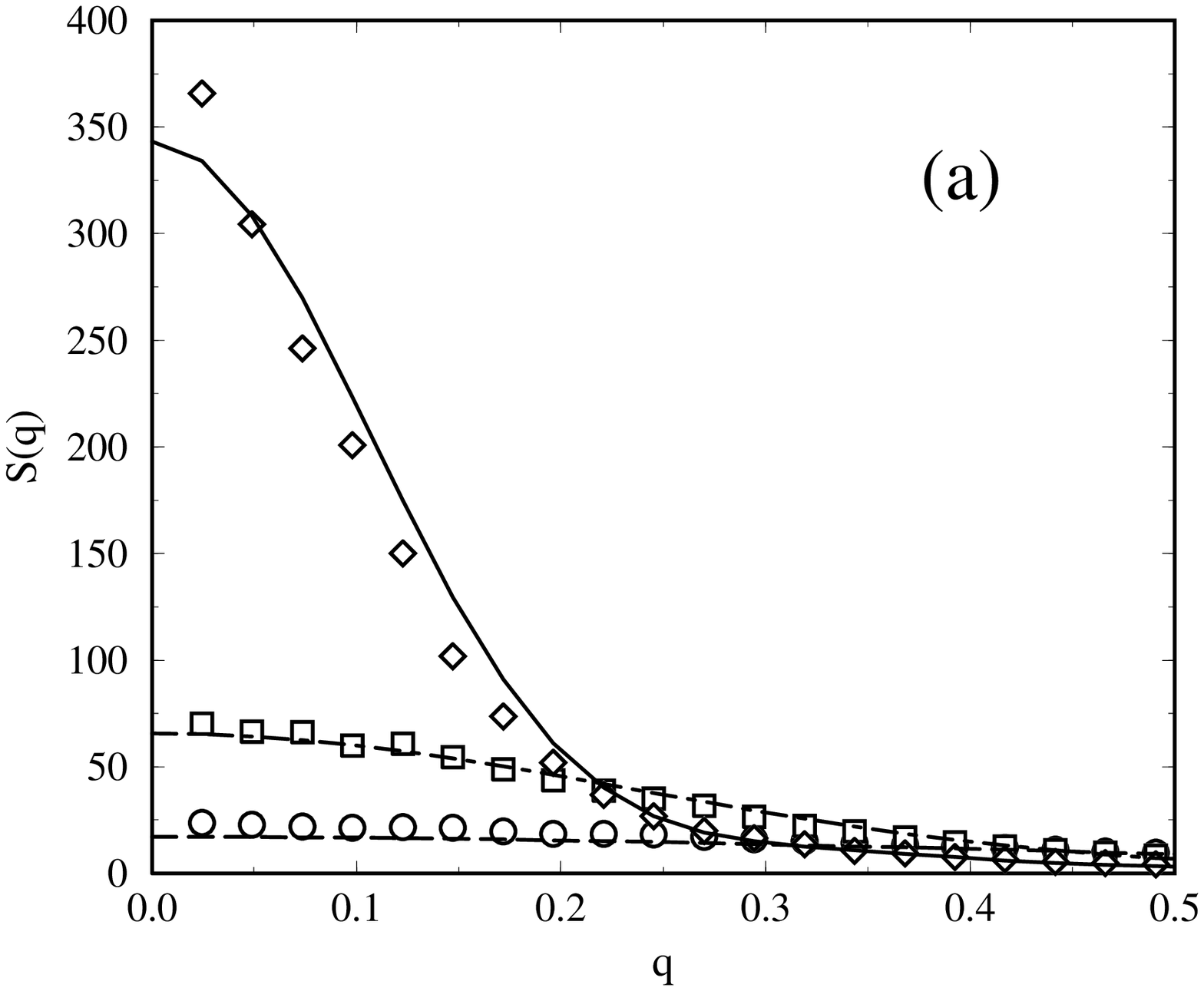}
\epsfxsize 3.5in \epsfbox{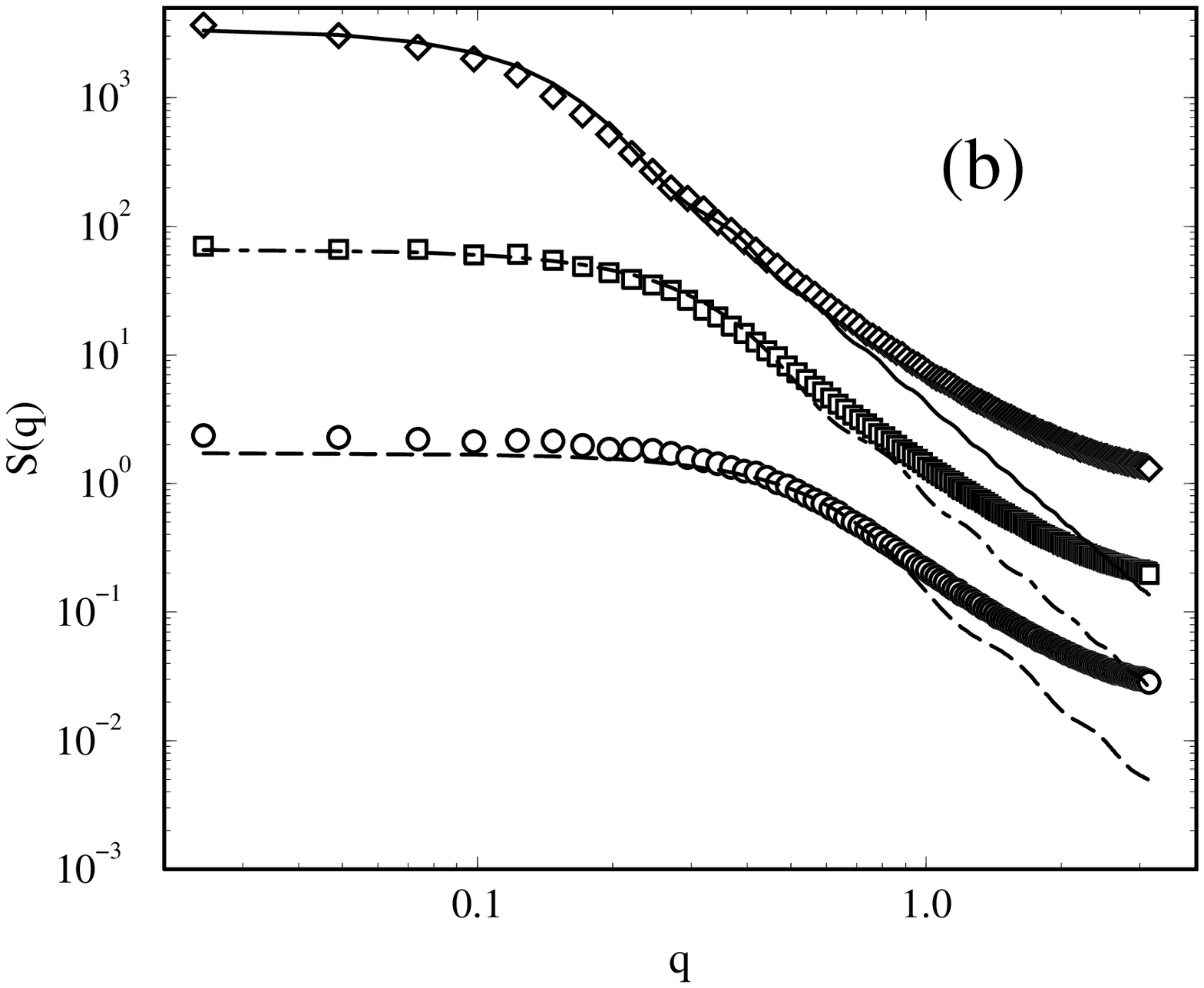}
\caption[]{
The structure factor
$S \left( q,\tau(|H|) \right)$.
The data points correspond to MC results and the curves to theoretical
results calculated by taking the Fourier transform of the
KJMA results for $G(r,t)$. Results are shown for
$|H|$=0.2 ($\Diamond$, solid curve), $|H|$=0.4 ($\Box$,
dotdashed curve), and $|H|$=0.8 ({\protect \large $\circ$},
dashed curve).
(a) Linear scale.
(b) Log-log scale.
{}For clarity, the results for $|H|$=0.2 and~0.8 have been offset by
$\pm$ two decades, respectively.
}
\label{fig_sq1}
\end{figure}

\begin{figure}
\includegraphics{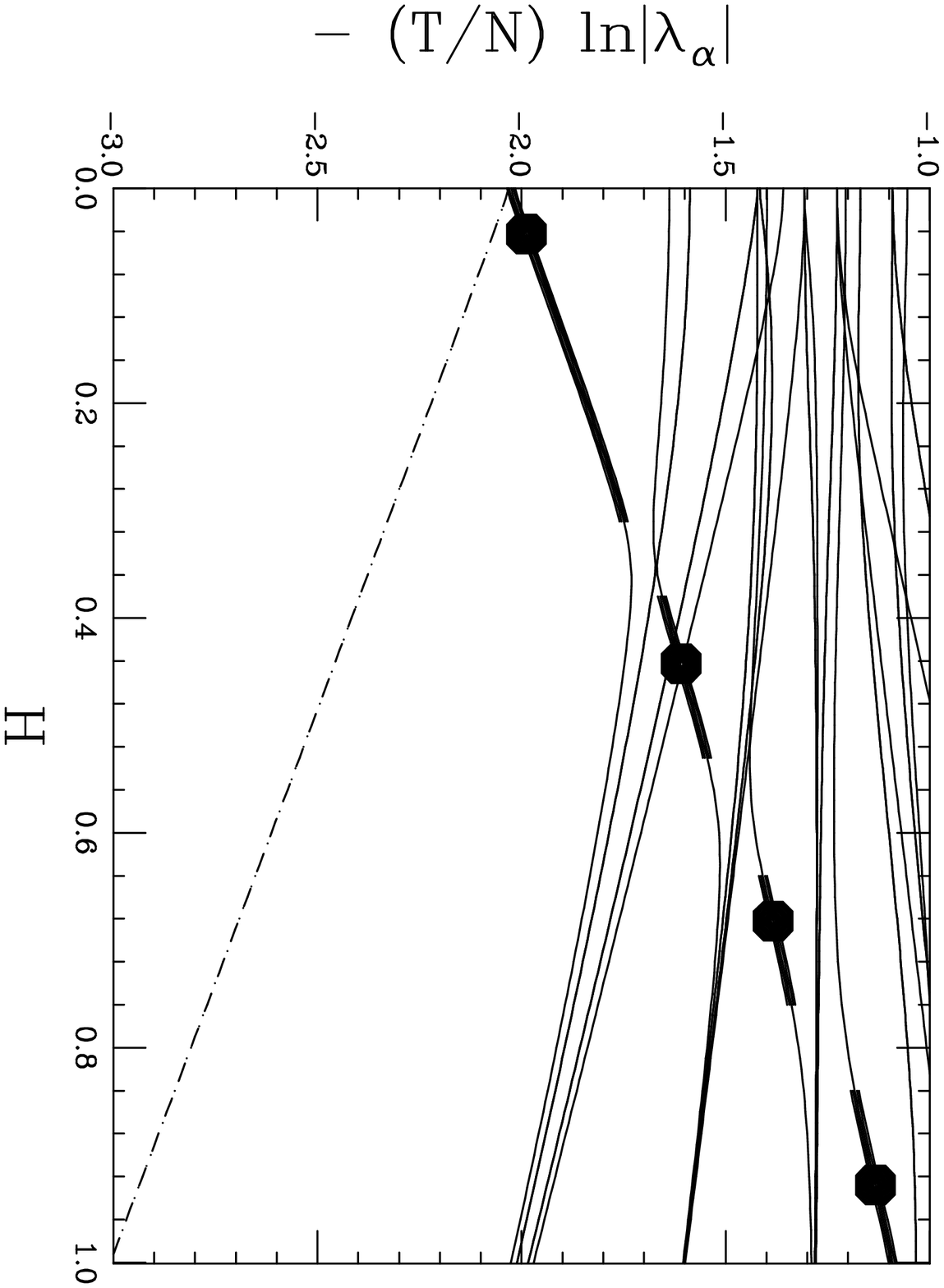}
\caption[]{
The nineteen largest 
transfer-matrix eigenvalues $\lambda_\alpha$, plotted vs $H$ 
as $-(T/N) \ln |\lambda_\alpha | $ for $N \! = \! 6$. 
The lowest-lying branch (dot-dashed line), which corresponds to 
the dominant eigenvalue $\lambda_0$, is the equilibrium free energy per spin.
The metastable branch is represented by the heavy curve segments. 
The solid circles represent the points along each segment, where the 
magnetization $m_\alpha(T,H)$ has its extremum. 
These points correspond to those similarly marked in 
Figs.~\protect\ref{fig_mh} and~\protect\ref{fig_lh}.
}
\label{fig_TMlambda}
\end{figure}

The analytic approximation used here 
for the propagation velocity of a field-driven
``tame'' interface in a two-dimensional, square-lattice
Ising model with Glauber dynamics will be 
described in detail elsewhere.\cite{RIKV99} It uses the SOS approximation 
for the equilibrium interface structure (which remarkably gives the 
exact surface tension for interfaces parallel to the lattice directions 
and an excellent approximation for inclined interfaces\cite{BURT51,AVRO82}) 
to estimate the populations in the different spin classes used in the 
$n$-fold way rejection-free Monte Carlo algorithm.\cite{BORT75} 
These class populations are used 
together with the contributions to the average propagation velocity from 
spins in each class, which are easily obtained from the transition 
rates, to calculate the overall average velocity. 
For the special case of Glauber transition rates, isotropic interactions, 
and an interface which is 
on average parallel to one of the symmetry directions of the lattice, 
the result is\cite{RIKV99} 
\begin{eqnarray}
\langle v(T,H) \rangle 
&=&
\frac{\tanh (\beta H)}{(1+X)^2} 
\Bigg\{
2 X 
+ 
\frac{1+X^2}{1 + 
\left(\frac{\sinh (2 \beta J)}{\cosh (\beta H)}\right)^2} 
\nonumber\\
& & 
+
\frac{X^2}{1-X^2}
\left[
X^2 
+ 
\frac{2 (1+2X)}{1 + 
\left(\frac{\sinh (2 \beta J)}{\cosh (\beta H)}\right)^2} 
\right]
\Bigg\} 
\;.  
\label{eq:vSOS}
\end{eqnarray}
Here $X = \exp (-2 \beta J)$ corresponds to a linear-response like 
approximation, in which the average class populations are approximated by 
their equilibrium values at $H \! = \! 0$, 
whereas $X = \exp (-2 \beta J) \cosh(\beta H)$ 
yields a nonlinear-response 
approximation which accounts for effects of the applied field 
on the nonequilibrium class populations. 
In Fig.~\ref{fig_vh}, $\langle v(T,H) \rangle$ from Eq.~(\ref{eq:vSOS}) is
shown vs $|H|$ at $T \! = \! 0.8T_{\rm c}$. The agreement of the 
nonlinear-response result with the directly
simulated ``tame'' interface velocities is excellent. 
Since the surface tension at 0.8$T_c$ is very close to isotropic, 
results for inclined interfaces are not needed here.




\begin{thebibliography}{100}
\bibitem{SHNEtoo} This method is also used in 
                  Ref.~\protect\onlinecite{SHNE98}.
\end{thebibliography}

\clearpage


\vspace*{10.0truecm}

\clearpage

\clearpage

\clearpage

\clearpage

\clearpage

\vspace*{8.0truecm}

\end{document}